%% file: ms.tex
\documentclass[twocolumn]{aastex63}
\usepackage[varg]{txfonts}
\usepackage{amssymb}
\usepackage{xspace}
\usepackage{natbib}
\bibpunct{(}{)}{;}{a}{}{,}
\usepackage{color}
\usepackage[normalem]{ulem}
\usepackage{color}
\usepackage{enumerate}
\usepackage{slantsc}
\newif\ifauth
\authtrue
\authfalse
\newcommand{\skipthis}[1]{}

\shorttitle{Multi-directional mass-accretion and collimated outflows in W51}
\shortauthors{Goddi, Ginsburg, et al.}

\begin{document}
\title{
Multi-directional mass-accretion and collimated outflows on scales 100-2000~au in early stages of high-mass protostars
}
\correspondingauthor{Ciriaco Goddi}
\email{c.goddi@astro.ru.nl,cgoddi@strw.leidenuniv.nl}


\author[0000-0002-2542-7743]{C. Goddi}
\affiliation{Department of Astrophysics/IMAPP, Radboud University Nijmegen, PO Box 9010, NL-6500 GL Nijmegen, the Netherlands}
\affiliation{ALLEGRO/Leiden Observatory, Leiden University, PO Box 9513, NL-2300 RA Leiden, the Netherlands}

  \author{A. Ginsburg}
\affiliation{Department of Astronomy, University of Florida, P.O. Box 112055, Gainesville, FL, USA}
  
\author{L. T. Maud}
\affiliation{European Southern Observatory, Karl-Schwarzschild-Str. 2, 85748, Garching bei M\"{u}nchen, Germany}         

\author{Q. Zhang}
\affiliation{Center for Astrophysics $|$ Harvard \& Smithsonian, 60 Garden Street, Cambridge, MA 02138}

\author{Luis A. Zapata}
\affiliation{Instituto de Radioastronom\'{i}a y Astrofis\'{i}ca, UNAM, Morelia, Michoac\'{a}n, M\'{e}xico, C.P. 58089}


\input{commands}

\begin{abstract}
We observed the W51 high-mass star-forming complex with ALMA's longest-baseline  configurations, achieving an angular resolution
of $\sim$20~milliarcseconds, corresponding to a linear resolution of $\sim$100~au at $D_{\mathrm{W51}}=5.4$~kpc. 
The observed region contains three high-mass protostars in which the dust continuum emission at 1.3~mm 
is optically-thick up to a radius $\lesssim$1000~au and has brightness temperatures $\gtrsim$200~K. 
 The high luminosity ($\gtrsim10^4$~\ls) in the absence of free-free emission suggests the presence of 
massive stars ($M\gtrsim20$~\ms) at the earliest stages of their formation. 
Our continuum images reveal remarkably complex and filamentary structures arising from compact cores. 
Molecular emission shows no clear signs of rotation nor infall on scales from 150 to 2000 au: we do not detect disks. 
The central sources drive young ($t_{\rm dyn}\sim$100~years),
fast ($v\sim$100~\kms), powerful ($\dot{M}>10^{-4}$~\msyr), collimated  outflows. 
These outflows provide indirect evidence of accretion disks on scales $r\lesssim$100--500~au (depending on the object).
The active outflows are connected to fossil flows that have different orientations on larger spatial scales, implying that the orientations of these small disks change over time. 
These results together support a variant of an accretion model for high-mass star formation in which 
massive protostars do not form a large, stable Keplerian disk during their early stages, but instead they
accrete material from multiple massive flows with different angular momentum vectors. 
This scenario therefore contrasts with the simplified classic paradigm of a  stable disk+jet system, which is the standard model for low-mass star formation, 
 and provides an experimental confirmation of a multi-directional and unsteady accretion model for massive star formation.
\end{abstract}


\section{Introduction}

The process by which  high-mass young stellar objects (HMYSOs) accrete their  mass is poorly understood.
One fundamental issue is that, for HMYSOs, the timescale for gravitational contraction (the Kelvin-Helmholtz time, $t_{\rm KH} = 10^3$ - $10^5$ years) is shorter than the timescale for accretion.
A consequent difficulty is that the star ignites nuclear burning reactions during its main accretion phase, which may  turn on powerful feedback processes (such as radiation pressure, ionising radiation, stellar winds, and outflows) that can halt or substantially reduce accretion.
These processes are either absent or insignificant for low-mass stars.
Different models have suggested that channeling material through a circumstellar accretion disk can overcome these feedback mechanisms \citep{Krumholz2007,Krumholz2009, Kuiper2010, Kuiper2011,Seifried2011,Klassen2016}.
Despite the theoretical support, the observational evidence of disks around OB-type protostars was limited \citep[e.g.,][]{BdW16,Cesaroni2017} 
and, as a consequence, their existence was still a matter of debate until very recently. 
 
Infrared (IR) interferometry  has revealed disks on scales of less than 1000 au around a limited number  of  infrared-bright HMYSOs  \citep[e.g.,][]{Kraus2010, Kraus2017,Boley2013,Frost2019}.
These HMYSOs are however relatively evolved and have apparently accreted most of their mass. Furthermore,  IR interferometry data alone  cannot  constrain the total extent of those disks, which requires observations at longer-wavelengths \citep[e.g.,][]{BdW16}. 
In this context,  
observations with (sub)millimetre (mm) interferometers allow access to  more embedded (i.e., less evolved) HMYSOs,  
and have led in recent years to the identification of about a dozen disk candidates with Keplerian signatures around luminous HMYSOs  (we report a full list of the disk sources, their properties, and relevant references in Appendix~\ref{app:HMdisks} and Table~\ref{tab:HMdisks}).
Among the clearest cases, we mention the archetypical B-type protostar IRAS 20126+4104 \citep[][]{Cesaroni2014}, the closest-known HMYSO Orion Source I \citep[][]{Ginsburg2018}, and the YSO G16.59$-$0.05  \citep[][]{Moscadelli2019}. 
In particular, with the advent of the Atacama Large Millimeter Array (ALMA) with its longest baselines, 
 the evidence for Keplerian-like disks around O-type HMYSOs has been increasing,
 including 
G17.64+0.16 \citep[][]{Maud2018,Maud2019}, G11.92-0.61 \citep[][]{Ilee2018},
G023.01$-$00.41 \citep[][]{Sanna2019},
and IRAS 16547$-$4247 \citep[][]{Zapata2019}.  Besides interferometric observations of mm thermal lines,
Very Long Baseline Interferometry (VLBI) studies have revealed that maser emission lines trace accretion structures and disk winds in the circumstellar environments of HMYSOs \citep[e.g.,][]{Matthews2010,Sanna2010a, Goddi2011b, MoscadelliGoddi2014}. 

Despite the growing evidence on both the theoretical and observational sides, it is still somewhat unclear whether the dozen identified disk-like structures are similar to their low-mass counterparts, dominated by the central YSO and in Keplerian rotation, or are self-gravitating, non-equilibrium   rotating entities.  The two structures may also coexist as a stable inner Keplerian disk embedded in a larger self-gravitating rotating envelope. 
In principle, even more complex pictures are possible for HMYSOs. 
For instance, recent 3D (magneto)hydrodynamic (MHD)  simulations of collapsing high-mass gas cores have shown that accretion on 1000~au scales is through filaments rather than a large disk \citep{Commercon2011,Myers2013,Seifried2015}. 
These filaments act as accretion streams which feed smaller Keplerian disks which are allowed to only form on scales of order $r\lesssim100$ au \citep{Myers2013, Joos2013, Seifried2012, Seifried2013, Seifried2015, Rosen2016,Rosen2019}. 
This mechanism allows the star to sustain high average accretion rates (of order of $10^{-3}$ \msyr) while mitigating the  effects of stellar feedback. 
This theoretical prediction has never been confirmed via observations.  
The fundamental question of how the highest-mass proto-O-stars  gather their mass remains therefore open. 
To test these exciting predictions on the formation mechanism for the most massive stars, observations with spatial resolution of order of 100 au  targeting pre-main-sequence HMYSOs  that are vigorously accreting are required.

With this in mind, we have used ALMA to study the high-mass star forming complex W51 at 1.3~mm with its longest baselines of 16 km. 
The W51 complex \citep[D$\sim$5.4 kpc;][]{Sato2010} is among the most luminous star forming regions in the Galaxy \citep[L$\sim 2\times 10^7$~\ls\footnote{This  luminosity is estimated using  Herschel Hi-Gal within a 2 pc radius, which includes both the W51 IRS2 and W51 Main protoclusters (see \citealt{Ginsburg2016} for details).};][]{Ginsburg2016} 
and is forming a very massive protocluster \citep[$>10^4$~\ms;][]{Ginsburg2016}, with over a hundred protostellar sources detected  \citep{Ginsburg2017}.  
The protocluster  contains both exposed O-type stars \citep{Rivera2020} and deeply-embedded  HMYSOs \citep{Ginsburg2017}, and as such it is a powerful laboratory for studying the entire high-mass star formation process. 

\begin{figure*}
\centering
\includegraphics[width=0.329\textwidth,trim={2.5cm 0cm 2.5cm 0cm},clip]{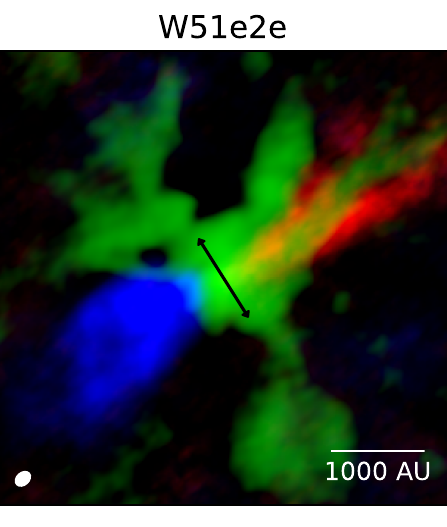}
\includegraphics[width=0.329\textwidth,trim={2.5cm 0cm 2.5cm 0cm},clip]{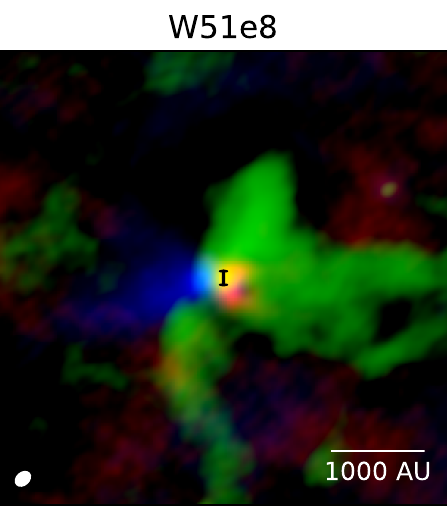}
\includegraphics[width=0.329\textwidth,trim={2.5cm 0cm 2.5cm 0cm},clip]{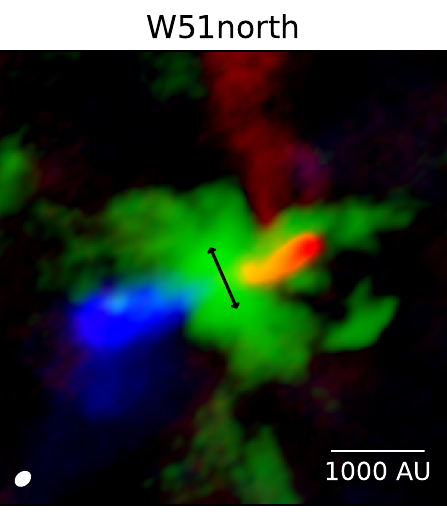}
\caption{Overlays showing the complex structures in the circumstellar material around the  high-mass protostars W51e2e, W51e8, and W51north (from left to right).  
The three-color images show the dust continuum emission at $\lambda$1.3~mm (in green),  tracing warm dust (50-400~K) in the (accreting) cores,  and  the redshifted and blueshifted emission of  the SiO J = 5-4 line (in red and blue), tracing  fast gas ($\pm100$~\kms) outflowing material from the same cores.   An area of $5000 \times 5000$~au is plotted in each panel, centered on 
 $\alpha(J2000) = 19^h 23^m 43^s.9659$,   $\delta(J2000) = +14^{\circ} 30' 34$\pas499 (W51e2e), 
 $\alpha(J2000) = 19^h 23^m 43^s.904$,   $\delta(J2000) = +14^{\circ} 30' 28$\pas244 (W51e8), and 
$\alpha(J2000) = 19^h 23^m 40^s.051$,   $\delta(J2000) = +14^{\circ} 31' 05$\pas483 (W51north), respectively 
(see Figures~\ref{figapp:outflow_e2e} and~\ref{figapp:outflow_north} for absolute coordinates). 
The black arrows indicate the orientation and the upper limits to the size of putative underlying disks.
The angular resolution of the continuum emission data is given by the ellipse drawn in the lower left corners (0\pas033 $\times$ 0\pas024 or 178 au $\times$ 130 au).  
}
\label{fig:rgb_overlays}
\end{figure*}

Here, we focus  on three HMYSOs: W51e2e, W51e8, and W51north, which are distributed across a $\sim3$ pc region.  
Previous observational studies using the Jansky Very Large Array \citep[JVLA;][]{Goddi2015b,Goddi2016,Ginsburg2016} and ALMA cycle 2  \citep{Ginsburg2017} have provided kinematic and physical properties of these HMYSOs on angular scales of 0\pas2--1\pas0 (corresponding to 1000-5000~au). 
These authors have also ruled out the presence of an observable ionized (\hii) region toward each of these targets down to a low ionizing continuum luminosity limit, indicating that  the central sources have not yet started ionising their surroundings via ultraviolet radiation and therefore must be in the earliest stages of their evolution. 
The longest baselines at 1.3~mm provide an angular resolution of 20--30 milli-arcseconds (mas), 
which allows us to unveil complex structures in the circumstellar material as well as collimated outflows at radii 100--2000 au, showcased in Figure~\ref{fig:rgb_overlays}, 
and thus to investigate the mass-accretion mechanism on the relevant scales to test theoretical models.

This paper is structured as follows.
Sect.~\ref{sect:observations}  summarizes the observations, data reduction, and imaging  of both the 1.3~mm continuum emission and selected molecular lines tracing dense gas and the outflows. 
Section~\ref{sect:analysis}    describes the data analysis aimed at deriving the physical (masses, temperatures, and luminosities) and kinematic (rotation, infall, and outflow) properties of the observed HMYSOs. 
Section~\ref{sect:disc}  highlights the main results  on the accretion/outflow structures identified in our high-angular resolution images. 
In Section~\ref{implications} we discuss the implications of these measurements in the context of our current theoretical understanding of high-mass star formation.
Conclusions are drawn in Section~\ref{sect:conclusions}.

\section{Observations, Data Reduction, and Imaging}
\label{sect:observations}

\subsection{Observations}
As part of the ALMA Cycle 3 program 2015.1.01596.S, we observed two fields centered on W51north [$\alpha(J2000) = 19^h 23^m 40^s.05$,   $\delta(J2000) = +14^{\circ} 31' 05$\pas5] and W51e2/e8 [$\alpha(J2000) = 19^h 23^m 43^s.91$,   $\delta(J2000) = +14^{\circ} 30' 34$\pas6] with  long baselines in Band 6 (216--237 GHz).
The project was carried out in three executions on 2015 October 27 and 30.  
Between 37 and 42 antennas in the 12-m array were employed and provided baselines ranging from 85 m to 16,196 m. 
The precipitable water vapour was between 0.85 and 1.98\,mm for the observations and there was reasonable phase stability (see below).  
We employed the band 6 sideband-separating receivers in dual polarization mode to record nine spectral windows (SPW). 
Seven SPWs had a $\sim$234 MHz bandwidth, were recorded with 960 channels and were Hanning smoothed by a factor of 2,
achieving a frequency resolution of 564~kHz (corresponding to $\sim$0.75 \kms~ at band 6). This setup enabled us to cover a large number of spectral lines from different molecular species.
 The remaining two SPWs had a broader bandwidth of 1.8 GHz and 1.13~MHz frequency resolution (corresponding to $\sim$1.5 \kms) to obtain sensitive continuum measurements at 217 GHz and 235 GHz as well as to cover additional spectral lines.  
We spent 2.2 hours on-source during a total observing time of 5.3 hours.

\subsection{Data Reduction and Imaging}  
The data calibration and imaging was carried out in the Common Astronomy Software Applications ({\sc casa}) package  (version 4.5.1)
and followed standard procedure. 
Fast referencing was used for these long baseline observations: the phase calibrator was observed every $\sim$70\,s.
The phase-calibrator to target-field separation angle was small at $<$1.2\,$^{\circ}$.
The combination of fast switching times and a close phase calibrator is imperative for accurate phase calibration interpolation for these long baselines.
Additionally, the water-vapour radiometer (WVR) scaling algorithm
\citep{Maud2017} was implemented to improve the short term (3-6\,sec) phase stability. This is a stand alone modular package that runs inside the {\sc casa} environment and provides a user with the optimal value to scale the  water vapour corrections in the \texttt{WVRGCAL} task within {\sc casa}.
The phase calibration of the data was then further improved by phase self-calibrating on the continuum emission down to an integration timescale of 50\,sec (using the {\sc casa} tasks \texttt{tclean} and \texttt{gaincal}).
The combined effect of WVR scaling and phase self-calibration improved the dynamic range of the final continuum maps by about 35\%.
The self-calibration solutions from the continuum were also applied to the full spectral dataset (using the {\sc casa} task \texttt{applycal}) in order to improve the signal-to-noise ratio of line cubes.

The calibrated visibilities were transformed from the Fourier plane ({\it uv} domain) into the image domain using the \texttt{tclean} algorithm. 
Three sets of images were produced with using uniform, Briggs (robust parameter R=0.5), and natural weighting schemes which achieved the angular resolutions of $\sim$19, 28, and 35 mas, respectively. 
Since interferometric observations lead to spatial filtering of large-scale structures,  the maximum recoverable scale in our images is 0\pas4  (about 2000 au at a distance of 5.4 kpc). This implies that we do not detect any emission that is extended over scales larger than 2000 au. In our images we only include data from baselines longer than 300 meters to mitigate striping in the images from large-scale emission detected on the shortest baseline that could not be properly imaged (baselines with a length below this value being predominantly in one direction).

\subsubsection{Continuum emission}
Continuum images were produced by selecting line-free channels in each SPW and then combining all SPW, and have a resulting central frequency of about 226 GHz assuming a flat-spectrum source.
Channel selection was accomplished in the {\it uv} domain using the shorter baselines $<$1000\,m where emission is seen and line-free regions can be discerned. Approximately 5 percent of the total  bandwidth was assessed to be `line-free' and was selected to establish the continuum level. 
 The final images were cleaned to a threshold of 0.3 mJy and then corrected for the primary beam response. 
 The lowest noise level in the images, away from bright sources, is typically $\sim0.1$~mJy~beam$^{-1}$, near the thermal noise level, as expected from the ALMA sensitivity calculator (near the brightest sources W51e2e and W51north, the noise can be two to three times higher).  
 Continuum images were produced with both uniform weighting and Briggs weighting  with a
robust parameter $R=0.5$, achieving resolutions of $\sim$19 and 28\,mas respectively. The former provides the highest angular resolution  to identify compact components, 
while the latter provides comparably better signal-to-noise ratio to recover the filamentary structure around the dusty peaks (displayed in Fig.~\ref{fig:rgb_overlays}).

\subsubsection{Line emission}
We produced spectral image cubes of each SPW.   
 For the emission line analysis, we used natural weighting (also excluding baselines shorter than 300 m) in order  to provide the highest signal-to-noise ratio, yielding a typical angular resolution of 35~mas. 
Images were also produced using a Briggs robust=0.5 weighting
which were useful in identifying more compact components. 
The typical RMS noise level was $\sim$1~mJy/beam in line-free channels and $\sim$2 mJy/beam in channels with strong line emission. 

The kinematic and moment analysis was performed outside of {\sc casa}. 
Firstly, the median value over the full SPW was used to estimate and subtract the local continuum \citep{Sanchez-Monge2017}. 
Secondly, from the full-SPW cubes we extracted cubes of all the identified spectral lines, using the \texttt{Astropy} module \texttt{spectral-cube}.
Thirdly,  for selected lines, we created integrated intensity (0$^{th}$ moment), velocity (1$^{st}$moment), and velocity dispersion  (2$^{nd}$moment) maps.
The 1$^{st}$ and 2$^{nd}$ moment maps were constructed using an intensity threshold of 7~mJy ($\sim 3-4 \sigma$, depending on spectral channel/source), and the velocity extent was chosen to avoid contamination from nearby lines where possible.

\input{table_src_masses}

\section{Results and Analysis}  
\label{sect:analysis} 

\subsection{Dust continuum sources}

Figure~\ref{fig:dust_tb} shows the 1.3~mm continuum emission, in units of brightness temperature ($T_B$), tracing warm dust in the W51e2e, W51e8, and W51north hot-cores, respectively. 
The dust emission is resolved into complex structures and displays multiple components on scales of 100~au to 3000 au. 
In the following, we 
estimate  sizes  (\S~\ref{cont:size}) and masses  (\S~\ref{cont:mass}) of different components,  as well as  luminosities (\S~\ref{cont:lum}) of the three dusty sources.

\subsubsection{Sizes}
\label{cont:size}

\paragraph{W51e2e.}
The dust emission is centrally peaked  (at $T_{B,max}=575$ K). 
 Around the brightest central beam, the dust emission  exhibits a clear central core object that is circular with a radius of $r\approx500$ au
(visualised by the yellow contour at $T_B = 130$~K in Fig.~\ref{fig:dust_tb}).
In the larger area around W51e2e, there are four main filamentary  structures that extend out to $r\sim2700$ au (seen as the red contour at $T_B = 15$~K in Fig.~\ref{fig:dust_tb}).

\paragraph{W51e8.}
The dust emission is centrally peaked  (at $T_{B,max}=435$ K). Fitting a Gaussian model to the continuum emission peak in both the uniform and Briggs maps provides a consistent deconvolved FWHM size of 0\pas028 (150 au).
The structure surrounding the bright central source is less symmetric than W51e2e. 
A filament   extends from the peak   to 1200 au along  northwest (NW; captured by the yellow contour at $T_B = 152$~K  in Fig.~\ref{fig:dust_tb}).
There is also a more extended structure toward the southwest (SW), covering approximately a semicircle centered on the brightest central peak out to a maximum radius of 2500 au (seen as the red contour  at $T_B = 15$~K in Fig.~\ref{fig:dust_tb}).

\paragraph{W51north.}
At variance with W51e2e and W51e8, the dust emission in W51north does not contain a central bright  point source, but displays a  slightly-elongated peanut-shaped structure with homogenous brightness (shown in yellow contour at $T_B = 245$~K in Fig.~\ref{fig:dust_tb}). 
Fitting a Gaussian model  in both the uniform and Briggs maps  provides a consistent deconvolved FWHM  long axis of  0\pas130 (700 au).
The  larger surrounding structure (captured in the red contour   at $T_B = 39$~K in  Fig.~\ref{fig:dust_tb}) is contained within a 1200 au radius and is primarily elongated northeast (NE) - southwest. 

\subsubsection{Masses}
\label{cont:mass}

For the mass estimates, we make a distinction between  optically thick and optically thin emission.  
 
 We first assume that in the highest column density regions (i.e. toward the source centers), the sources are just barely optically thick ($\tau=1$). Therefore, $N(\hh) = \tau / \kappa_{\nu} = 1/\kappa_{\nu}$ where $\kappa_{\nu}$ is the dust opacity coefficient.
 We use the same coefficient $\kappa_{227\mathrm{~GHz}} = 0.0083 \mathrm{~cm}^2 \mathrm{g}^{-1}$
as in  \citet[][]{Ginsburg2017}, 
 resulting in  $N(\hh) = 2.6\ee{25}$ \persc.
The mass in a beam is then $M_{beam} = N(\hh) \times A = A / \kappa$, where $A$ is the beam area (0\pas033 $\times$0\pas024) in cm$^2$,  which yields dust masses  of each source $M_{Beam}=0.36$ \ms.

 Although the highest column density peak positions have increased optical depth, the emission surrounding the peak position is likely to be optically thin. 
We can then use the flux density measurements to compute gas masses using the following equation: 
$$M = \frac{S_{\nu} d^2}{\kappa_{\nu} B_{\nu} (T_d)},$$   
where $S_{\nu}$ is the source flux density, 
$d$ is the distance to the source, $\kappa_{\nu}$ is the dust opacity coefficient (which includes the gas-to-dust ratio of 100), $B_{\nu} (T_d)$ is the Planck function for a blackbody at dust temperature $T_d$. 
The latter is not known but we can estimate lower limits to $T_d$ in two different ways. 
The first estimate is provided by the peak brightness temperature of the mm continuum emission at the source center (where we assume that the emission is optically thick). 
The peak intensities of the three sources in the robust 0.5 maps are $S_{227\mathrm{~GHz}}(\mathrm{e2e}) = 19.4$ mJy \perbeam, $S_{227\mathrm{~GHz}}(\mathrm{e8}) = 14.7$ mJy \perbeam, and $S_{227\mathrm{~GHz}}(\mathrm{north}) = 13.2$ mJy \perbeam,
which correspond to brightness temperatures of $T_{B,max}=$575~K, 435~K, and 390~K, respectively.
These   provide the first lower limits on $T_d$. 
A second estimate comes from LTE modeling of CH$_3$OH lines imaged with the ALMA cycle 2 program \citep{Ginsburg2017}, which provided temperatures in the range 200-600~K inside 5000 au. Therefore we can assume $T=T_{gas, min} = 200$~K as a second lower limit on $T_d$.  

In the following we use these two approaches (assuming thin and thick emission) and  these two independent $T_d$ lower limits to obtain mass estimates of the spatially resolved cores.

\paragraph{W51e2e.}
The central core object  spans 27 beam areas (radius $r\approx500$ au), implying a mass $M_{e2e,core} = 0.36 \times 27 = 9.7$ \ms\ if it is entirely optically thick.
If it is optically thin on average, using the expression above for $T_d=T_{B,max}=575$ K, the total is 4 \ms, while for $T_d=T_{gas, min}=200$ K, the total is 12 \ms.  
In the surrounding filamentary  structures we estimate the mass assuming $T_d=200$ K and obtain $M_{e2e,filaments} = 25$ \ms (excluding the inner core).

\paragraph{W51e8.}
For the filament   extending  to 1200 au from the central peak  along  NW we measure a mass of 4~\ms  for $T_d=T_{B,max}=435$ K and  9~\ms for $T_d=T_{gas, min}=200$ K.
For extended structure toward the southwest, we obtain $M(200 K) = 16$ \ms.

\paragraph{W51north.}
The central peanut-shaped structure occupies 14 beams, implying a lower-limit mass $M_{north,core} = 0.36 \times 14 = 5$ \ms if it is optically thick or $M=4.2$ \ms if it is optically thin and isothermal at $T_d=T_{B,max}=390$ K.
For the larger surrounding structure  we obtain a mass range $M(390 \mathrm{~K}) = 14$ \ms and $M(200 \mathrm{~K}) = 28$ \ms.
 \\

 Table~\ref{tab:src_masses} summarizes the sizes and masses of the different components:  centrally-peaked core, surrounding compact core, and filaments.

We compare the recovered flux in the long-baseline maps to that in the ALMA cycle 2 maps at the same frequency   \citep{Ginsburg2017}.
In the robust 0.5 maps, in W51e2e, 13\% of the flux is recovered, in W51e8, 12\%, and in W51north, 23\%.
These low recovery fractions indicate that most of the dust emission is smooth on $\gtrsim2000$ au scales. 
It is therefore likely that the cores resides in a larger envelope of dense material containing up to five to ten times as much
mass within $r\lesssim 2000$ au as we have estimated above
\citep{Ginsburg2017}.

\subsubsection{Protostellar luminosities}
\label{cont:lum}
The peak brightness temperatures are a lower limit to the surface brightness of the millimeter core, since an optical depth $\tau < 1$ or a filling factor of the emission $ff < 1$  would both imply higher intrinsic temperatures. 
One can use these temperatures  to estimate lower limits on the source luminosities.  
Assuming blackbody emission from a spherical beam-filling source, the luminosity can be calculated as  $L = 4\pi r^2 \sigma_{sb} T^4$, where 
$\sigma_{sb} = 5.670373 \times 10^{-5} g s^{-3} K^{-4}$ is the Stefan-Boltzmann constant. 
This provides lower limits of $8.2 \times 10^3$~\ls, $7.5 \times 10^3$~\ls, and $2.8 \times 10^3$~\ls\ for the luminosity of W51e2e, W51e8, and W51north, respectively. 
Such luminosities correspond to B1-B1.5V main-sequence stars with masses $M\sim8-11$ \ms\ and effective temperatures of 24000-26000 K \citep{PecautMamajek2013}.  
However, the coarser-resolution observations carried out with ALMA in cycle 2  recovered more flux within a 0\pas2 radius, yielding  slightly higher luminosities, $L\gtrsim10^4$ \ls, implying masses $M\gtrsim10-15$ \ms \citep{Ginsburg2017}.  
Even these larger-scale measurements are luminosity
lower limits, as an unknown fraction of the luminosity escapes to much larger
radii along the outflow cavities without being reprocessed by dust.

\begin{figure*}
\centering
\includegraphics[width=0.329\textwidth]{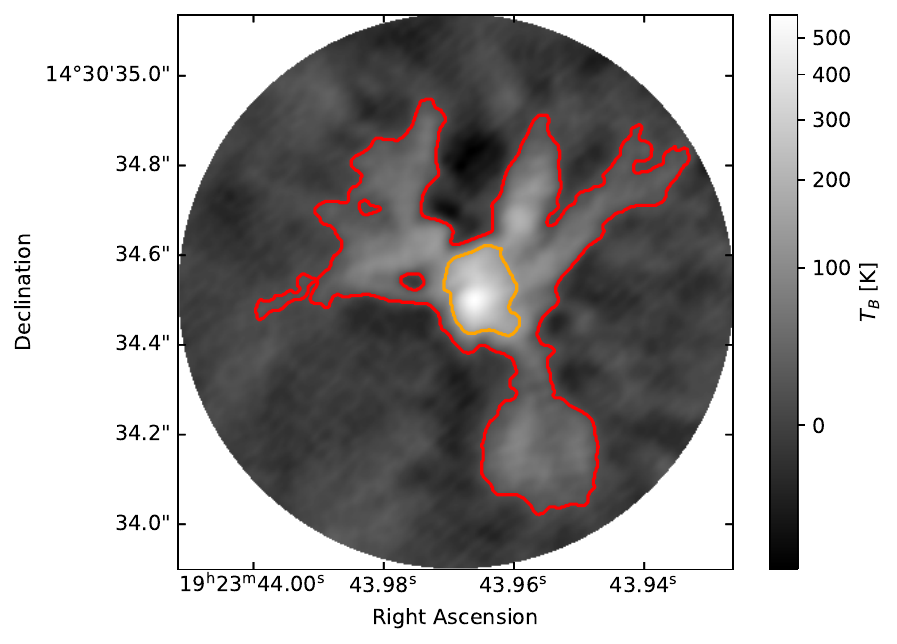}
\includegraphics[width=0.329\textwidth]{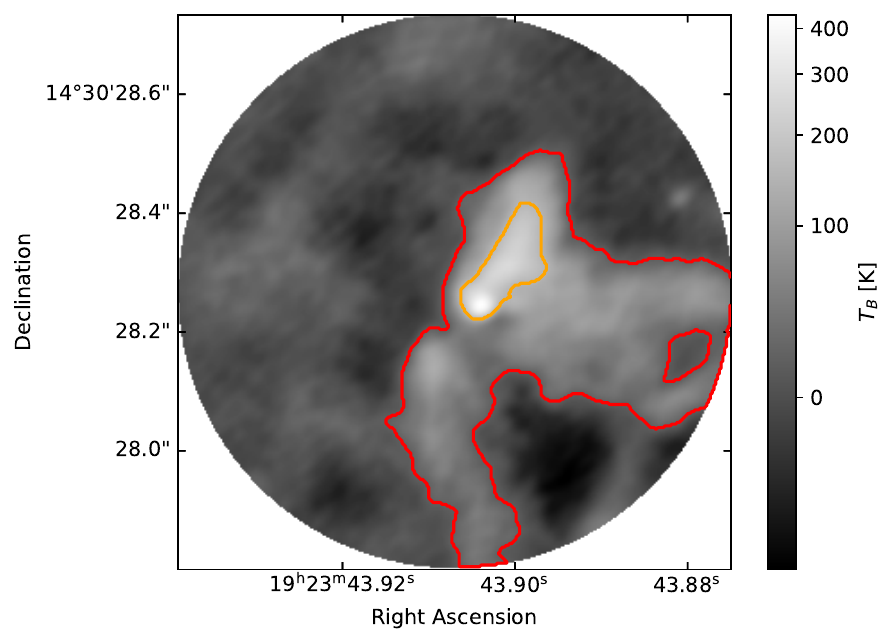}
\includegraphics[width=0.329\textwidth]{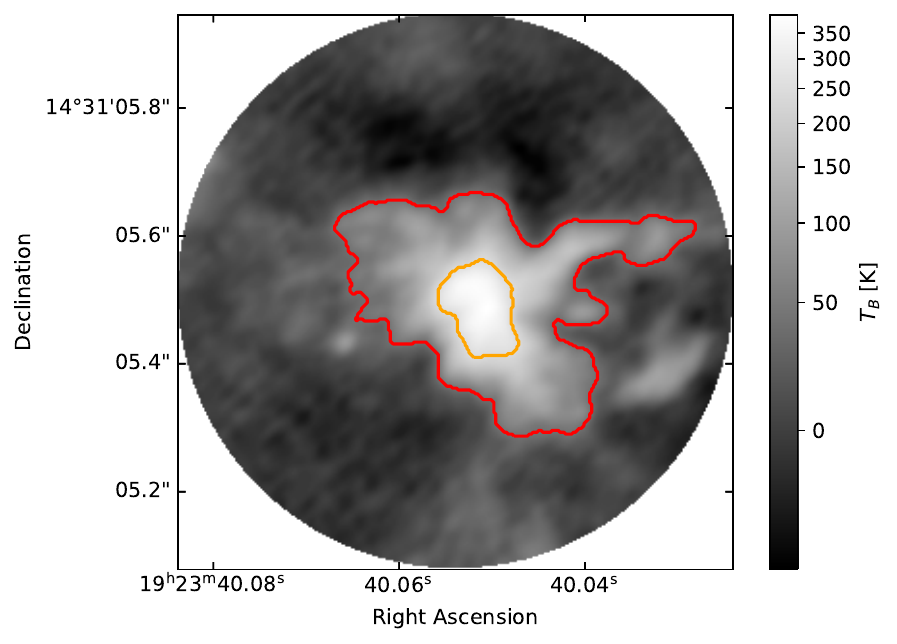}
\caption{
The greyscale shows the dust continuum emission,  tracing warm dust in the W51e2e, W51e8, and W51north hot-cores (from left to right). 
The contours enclose the regions where we compute the mass values quoted in the paper, for both  the central cores (yellow contours) as well as  the full extent of the continuum (red contours). 
In particular, the orange  and red contours trace the 23\% and 3\% (131 and 15 K), 35\% and 3\% (152 and 15 K), 62\% and 10\% (245 and 39 K) levels from the continuum flux-density peaks of 19, 14, and 13 mJy/beam, for W51e2e, W51e8, and W51north, respectively. The brightness temperature T$_B$ scale is on the right-hand side in each panel. 
}
\label{fig:dust_tb}
\end{figure*}
\begin{figure*}
\centering
\includegraphics[width=\textwidth]{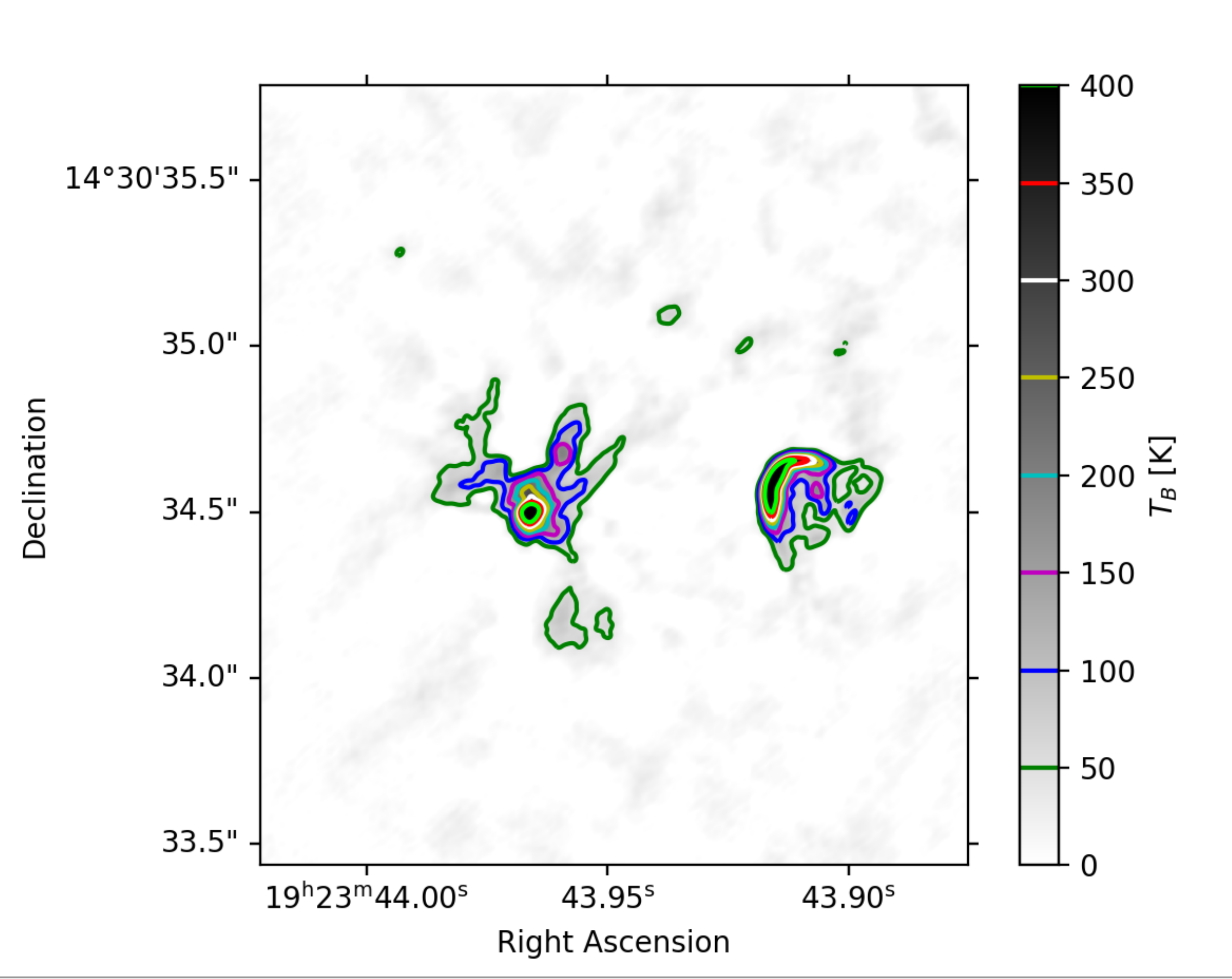}
\caption{
The greyscale shows the 1.3~mm continuum emission,  tracing warm dust in the W51e2 core. The dusty hot core e2e is to the left, and the hyper-compact \hii\,region e2w (not discussed in this article) is to the right.  
The contours enclose the regions with different brightness temperatures T$_B$ (scale on the right-hand side).
While the highest temperatures in e2e are measured towards the central core, the dusty streamers are also significantly warm with 50--150~K. 
}
\label{fig:dust_tb_e2}
\end{figure*}

\subsection{Kinematics of dense molecular gas}
\label{kinematics}

Each of the three target sources is a typical hot molecular core and is extremely rich in spectral lines \citep[see][]{Ginsburg2017}. 
Within the plethora of identified spectral lines, we  identified a subset of lines which are unblended and strong enough to be used for kinematic analysis. 
The molecular lines observed directly toward the dusty structures are only seen in absorption against the continuum, with the exception of SiO, which is seen in the outflow (see \S~\ref{res:outflows}).

\begin{figure*}
\includegraphics[width=\textwidth]{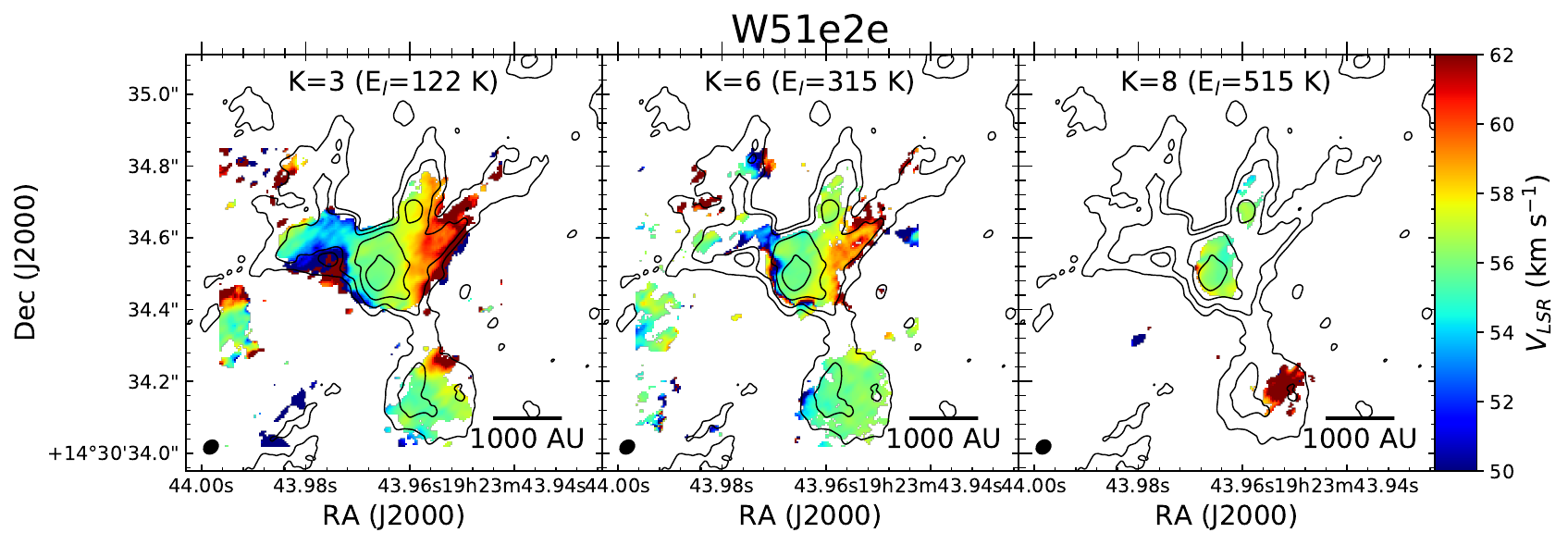}
\includegraphics[width=\textwidth]{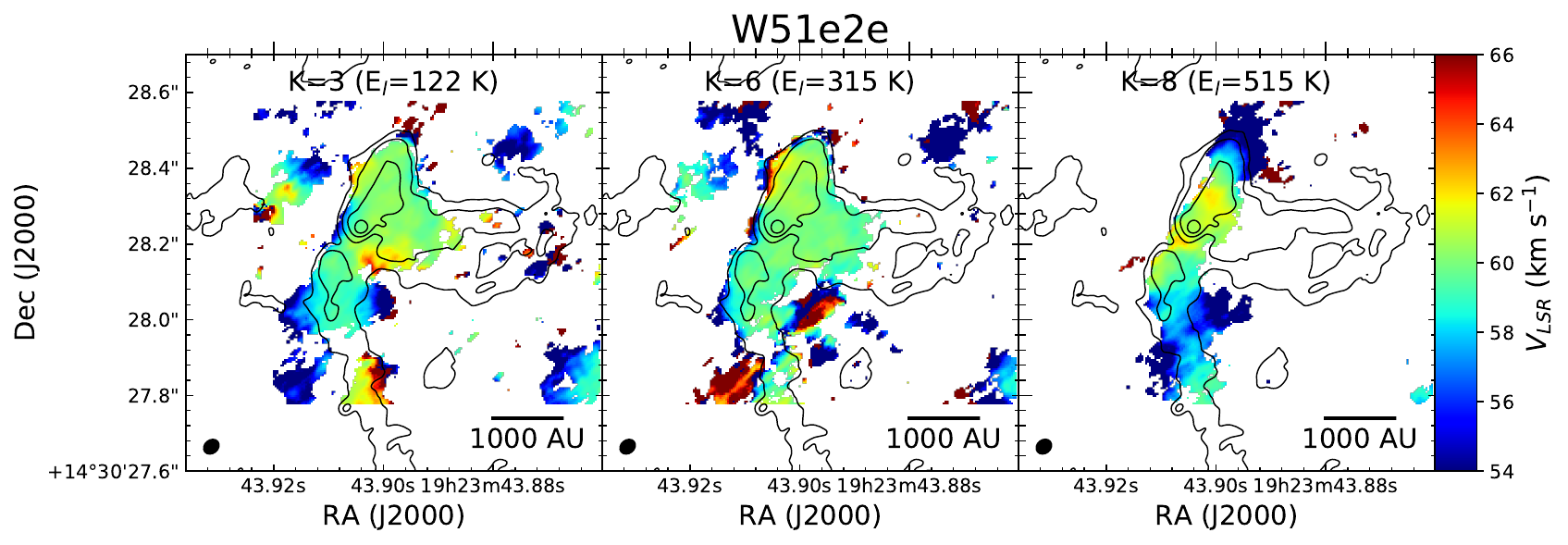}
\includegraphics[width=\textwidth]{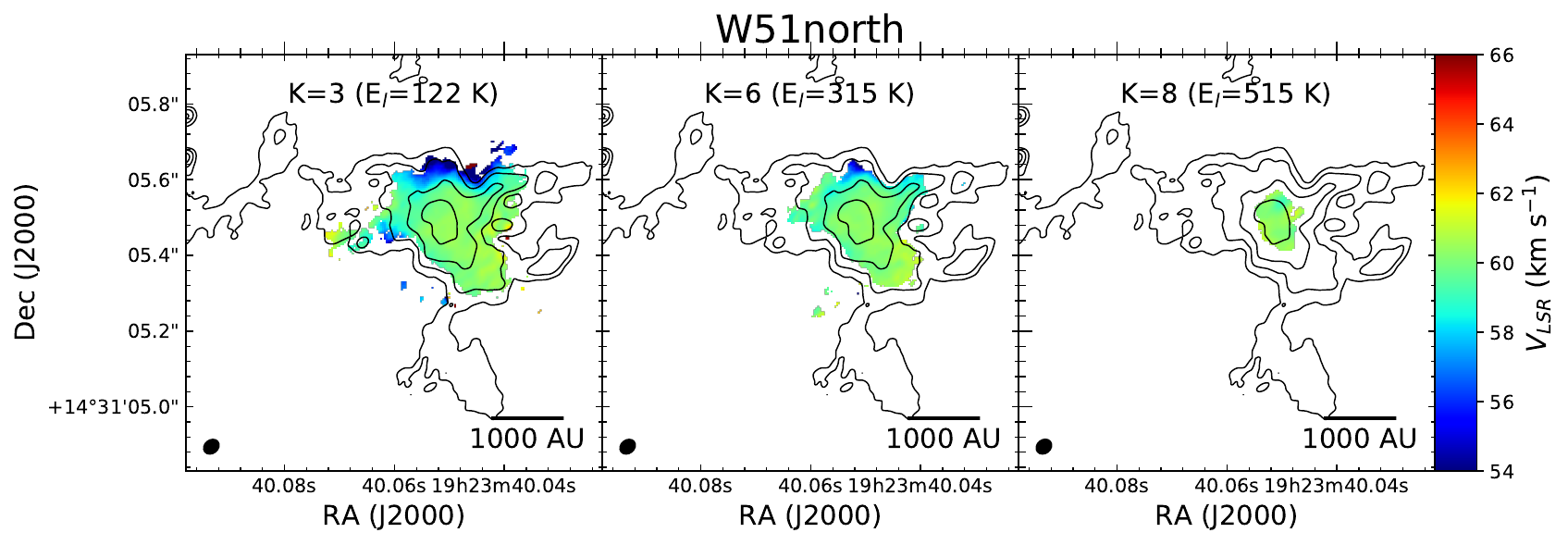}
\caption{ 
Velocity fields of the circumstellar molecular gas in W51e2e (top panels), W51e8 (middle panels), and W51north (bottom panels). 
Each row displays the $^{1st}$moment maps of the CH$_3$CN J=12-11 K=3 ($E_l$=122 K), K=6 ($E_l$=315 K), K=8 ($E_l$=515 K) transitions (colors) seen in absorption against the 1.3~mm continuum emission (shown in black contours). The LSR velocity scale is drawn on the right-hand side. 
The angular resolution of the velocity maps is given by the filled ellipse drawn in the lower left corners: 
 0\pas041 $\times$ 0\pas031 or 221~au $\times$ 167~au (the continuum maps achieve a better resolution: 0\pas033 $\times$ 0\pas024 or 178 au $\times$ 130 au). 
 \vsys\ are 56~\kms (W51e2e) and 60~\kms (W51e8, W51north), as measured from highly-excited ammonia inversion lines \citep{Goddi2015b,Goddi2016})
Note how the different CH$_3$CN  lines from the K-ladder, show exactly the same remarkably flat velocity profile  regardless of their excitation. 
The only exception is W51e2e which shows red-shifted velocities along the  NW dust lane (see \S~\ref{res:acc_fil} for an interpretation).
Transitions from other molecular species show similar flat profiles.  
}
\label{fig:mom1}
\end{figure*}

In order to  study gas kinematics, we used intensity-weighted velocity  maps and 
velocity dispersion   maps of different dense gas tracers, which achieve  a typical spatial resolution of about 150--200 au. 
In Figure~\ref{fig:mom1}, we display  velocity field maps  for selected lines of CH$_3$CN\footnote{CH$_3$CN is a typical dense gas tracer \citep[e.g.,][]{Cesaroni2017} and
is often seen to trace rotation in disk-like structures \citep{Ilee2018,Johnston2015}, although it is not always the case \citep{Maud2018}.}  for W51e2e, W51e8, and W51north, respectively. 
A remarkable feature seen in these velocity maps is the nearly perfectly flat velocity field in all three sources near their systemic velocity (shown in green). 
This property is best seen towards W51north, where different CH$_3$CN  lines from the K-ladder (with lower energy levels between 122~K and 515~K), show exactly the same remarkably  flat velocity profile regardless of their excitation (Fig.~\ref{fig:mom1}, bottom row). 
This result does not depend on the molecular tracers or the probed scales, and thus mirrors real physical properties of the targeted HMYSOs. 
The only exception is W51e2e (Fig.~\ref{fig:mom1}, top row, left panel), which shows red-shifted velocities along the  NW dust lane (see \S~\ref{res:acc_fil} for an interpretation).

In Appendix~\ref{app:lack_of_kin} we review possible explanations for the lack of velocity structure towards these cores  
and we conclude that it is an observational effect caused by the high optical depth in the dust continuum, as predicted in \citet{Krumholz2007}.

\subsection{Protostellar outflows}
\label{res:outflows}

We detected bright SiO ($J= 5-4 \ v=0$ line) emission 
which traces compact bipolar outflows stemming from the three HMYSOs. 
These small-scale SiO outflows are barely detected beyond $r \sim 2000$ au, but they connect to larger-scale outflows  (2000 au up to 30000~au) detected in $^{12}$CO ($J=2-1$)\footnote{\citet{Shi2010b} also mapped the outflow from W51e2e in the $^{12}$CO $J=3-2$ line.}  by \citet{Ginsburg2017}.
The outflows from W51e2e and W51north are shown in Figures \ref{fig:outflow_e2e} and \ref{fig:outflow_n} (see Appendix~\ref{app:outflows} for a detailed description of their  structures).

Both CO and SiO outflows display high speeds (with peak velocities $\pm100$ \kms).   
Such high speeds coupled with the compact sizes imply that these outflows are dynamically very young. 
In particular,  the CO outflows have  a dynamical age of a few thousand years, whereas the SiO  outflows have  a dynamical age of roughly one hundred years (see Appendix~\ref{app:outflow_age} for dynamical age estimates).

While the W51e2e outflow has a simple bipolar morphology, the W51north outflow shows a sharp change from small  to
large  scales.  The inner SiO outflow is simple and bipolar at position angle (P.A.) --70\dg\ out to $r<1300$ au, at which point the redshifted flow is sharply truncated (the blue-shifted lobe extends out to $r\lesssim 1600$ au).  
 Starting at about 800~au from the center, the redshifted SiO emission suddenly turns by 80\dg (at P.A.$\sim$10\dg) and 
 continues to the north out to $\sim 5000$ au, where it meets the large-scale $^{12}$CO outflow which extends out to $\sim 20000$ au at P.A. --13\dg.   The outflow is also detected in \wat\ maser emission, which traces the SiO morphology and kinematics consistently \citep[][Fig.~\ref{fig:outflow_n}, lower panel]{Imai2002}.

We discuss possible scenarios to explain this peculiar outflow structure in \S~\ref{disc:outflow_n}. 

\begin{figure*}
\centering
\includegraphics[width=0.45\textwidth]{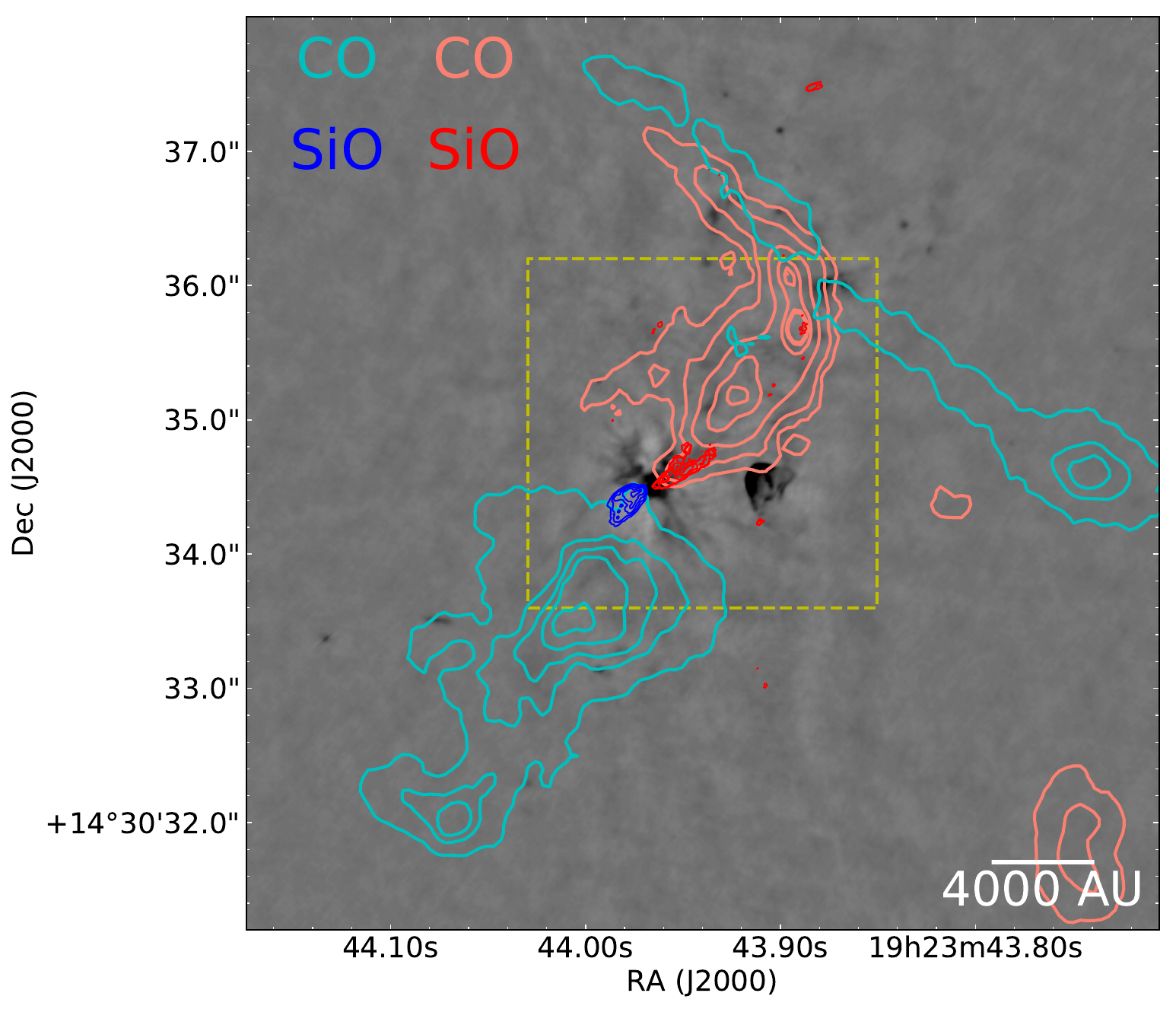}
\includegraphics[width=0.5\textwidth]{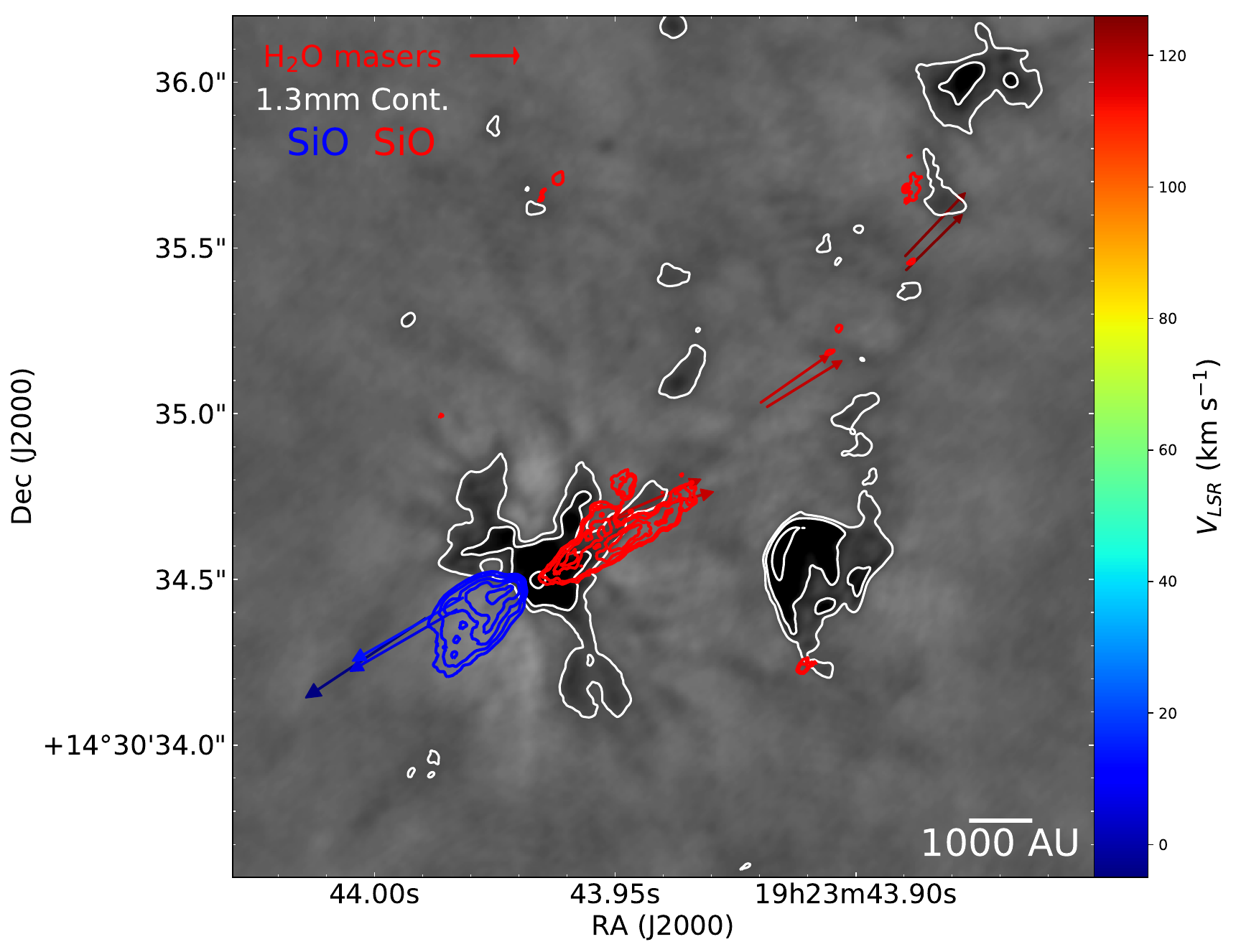}
\caption{ Outflow from W51e2e. 
({\it Left panel}) The outflow is traced by the \co\  $J=2-1$ line from ALMA cycle 2 (beamsize 0\pas2) and SiO $J=5-4$ line from ALMA cycle 3 (beamsize 0\pas035). 
The CO emission is integrated over [0,45]~\kms\  and [73,180]~\kms, for the blueshifted (cyan contours) and redshifted (salmon contours) emission, respectively.
The SiO emission is integrated over [-56,53]~\kms\ and [71,159]~\kms, for the blueshifted (blue contours) and redshifted (red contours) emission, respectively. 
These corresponds to V$_{\rm outflow}$ = [-112,-3]~\kms\ and [15,103]~\kms\ for \vsys=56~\kms.
The yellow dashed rectangle indicates the zoomed region plotted in the right panel. 
({\it Right panel}) The outflow  is traced by SiO emission and H$_2$O masers.
The  H$_2$O masers 3-D velocities measured with the VLBA \citep{Sato2010} are overplotted  onto the total intensity of the redshifted (red contours) and blueshifted (blue contours) emission of  the SiO J = 5-4 line and the $\lambda$1.3~mm continuum emission (white contours and greyscale image).
 Colors denote maser l.o.s. velocity (color scales on the right-hand side). 
The blueshifted and redshifted water masers closer to the protostar imply a flow P.A. of --56\dg, perfectly consistent with thermal SiO, while the redshifted masers along  the outflow at larger distances have a P.A. of --43\dg, closer to the \co\,outflow axis away from the protostar.}
\label{fig:outflow_e2e}
\end{figure*}

\begin{figure*}
\centering
\includegraphics[width=0.4\textwidth]{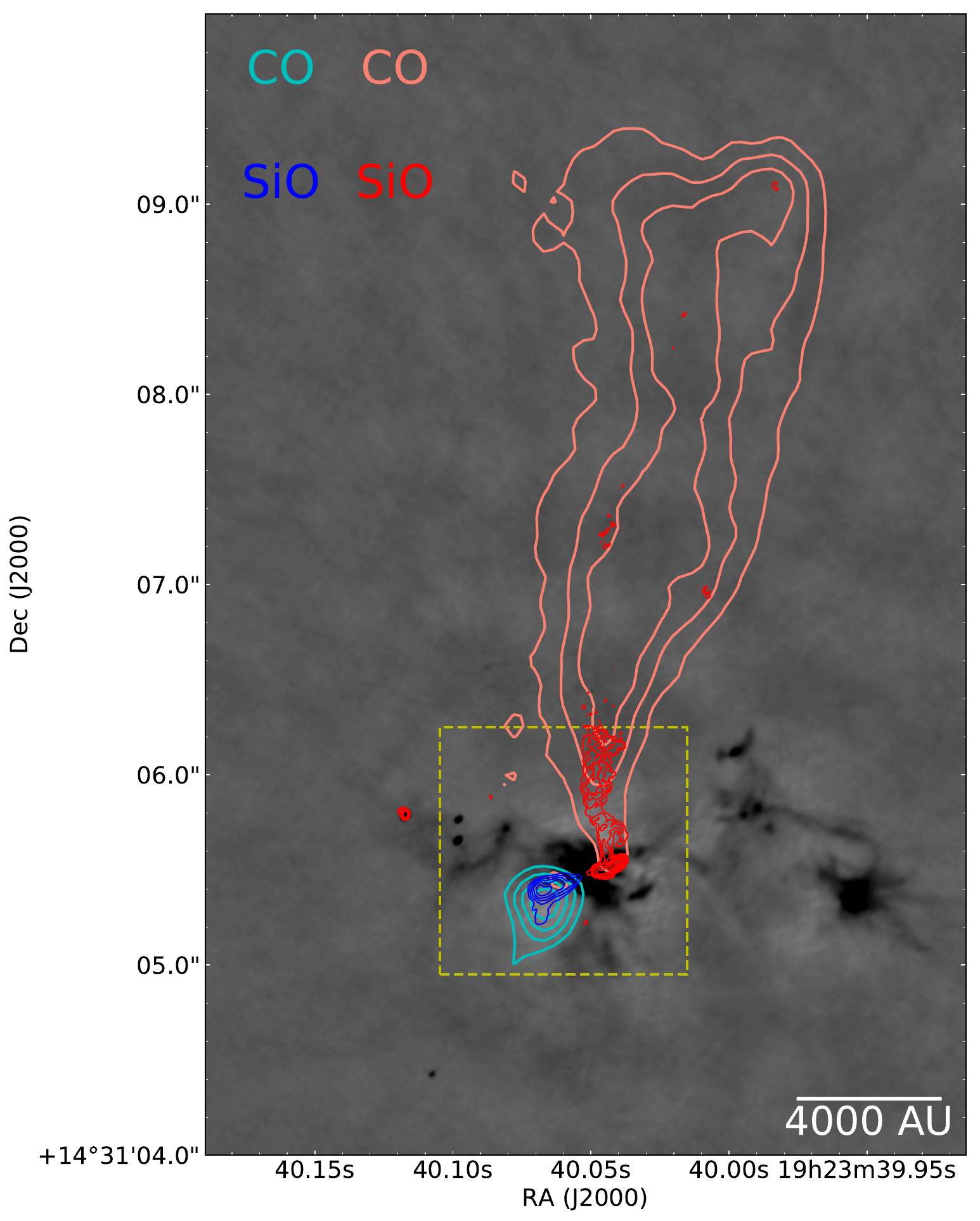}
\includegraphics[width=0.55\textwidth]{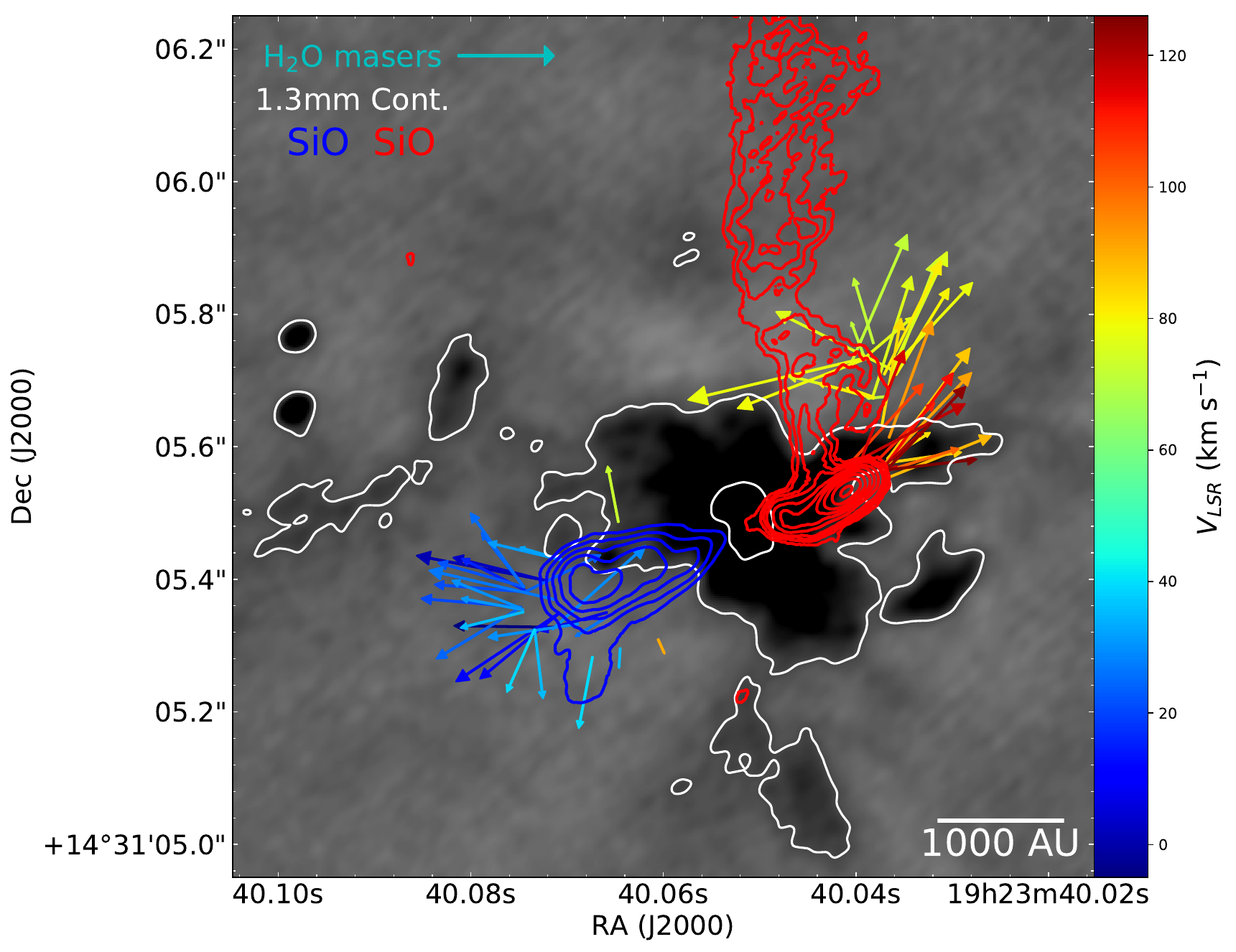}
\caption{ Outflow from W51north. 
({\it Left panel}) The outflow  is traced by the \co\  $J=2-1$ line from ALMA cycle 2 (beamsize 0\pas2) and SiO $J=5-4$ line from ALMA cycle 3 (beamsize 0\pas035). 
The SiO emission is integrated over [-36,56]~\kms\ and [66,126]~\kms, for the blueshifted (blue contours) and redshifted (blue contours) emission, respectively. These corresponds to V$_{\rm outflow}$ = [-96,-4]~\kms\ and [6,66]~\kms\ for \vsys=60~\kms. 
The yellow dashed rectangle indicates the zoomed region plotted in the right panel. 
({\it Right panel}) The outflow  is traced by SiO emission and H$_2$O masers.
The  H$_2$O masers 3-D velocities measured with the VLBA \citep{Imai2002} are overplotted  onto the total intensity of the redshifted (red contours) and blueshifted (blue contours) emission of  the SiO J = 5-4 line and the $\lambda$1.3~mm continuum emission (white contours and greyscale image).
 Colors denote maser l.o.s. velocity (color scales on the right-hand side). 
 The \wat\ masers reside in two complexes separated by $\sim$3000~au at the front edges of the compact SiO outflow lobes and are moving away from each other in ballistic motions at about 200~\kms\  and at a P.A. of~--72\dg, consistently with the SiO  NW- SE  outflow. 
 }
\label{fig:outflow_n}
\end{figure*}

\subsection{Momentum and ejection rates of outflows}
\label{analysis:momentum_rate}

We can use the SiO emission structure to estimate the momentum and ejection rates of the outflows. 
One straightforward method would be  to determine the mass of the outflowing gas from SiO emission and divide it by the outflow dynamical age. 
While this may provide the correct order-of-magnitude answer in some cases, molecular outflows, such as those seen in CO and SiO emission, are driven by an underlying primary wind/jet  and probe mostly entrained ambient gas moving at lower velocity (this applies especially to CO).

Since in protostellar jet/outflow systems momentum is conserved (we assume momentum-driven outflows; e.g., \citealt{MassonChernin1993}), one can set a tight constraint
on the mass loss rate using $\dot{m_j} v_j = \dot{m_o} v_o = \dot{P}_o$, where $\dot{m}$ is the mass-loss rate, $v$ is the speed and $\dot{P}$ is the momentum rate ($j$ and $o$ stand for the primary jet and molecular outflow, respectively).

The outflow momentum rate can be obtained by dividing 
 the outflow  momentum  by its  age. 
In  Appendix~\ref{app:outflow_age} we   
estimate the dynamical age of the outflows from the ratio of their  projected  length  to their maximum speed, by assuming an inclination $i=45$\dg.  
In Appendix~\ref{app:momentum-rate} we estimate the momentum of the outflowing gas  
by  computing  its  mass from the SiO emission integrated in the full velocity range. 
In order to convert from SiO column to \hh\ column, we need to fix the abundance of SiO, 
$X_{SiO} = N(SiO) / N(\hh)$.  
Towards a  sample of 14 high-mass star forming regions, \citet{Leurini2014} used the APEX single-dish telescope to target the SiO (J=3-2 and J=8-7) lines, obtaining $X_{SiO} =0.7-4.8\ee{-8}$  while \citet{Sanchez-Monge2013b} used the IRAM-30m telescope to target the same sample in the SiO (J=2-1 and J=5-4) lines obtaining $X_{SiO} =0.1-3\ee{-8}$.
We assume a
 conservative high abundance of SiO $X_{SiO} = 10^{-7}$ (a lower abundance of SiO would imply higher mass and momentum in the outflow). 
The ages, momenta, and the resulting momentum rates derived with this analysis are  reported in columns 5, 7, and 8 of Table~\ref{tab:outflows}, respectively. The momentum analysis was conducted on the blue-shifted lobes (which display a simpler structure; see Appendices~\ref{app:outflows} and \ref{app:momentum-rate}), therefore the total outflow mass and momentum rate  are found multiplying those numbers by a factor of 2 (to account for both outflow lobes), yielding ${M}_o = 0.72, 0.36,  0.22$~\ms\ and $\dot{P}_o =0.42, 0.24,  0.06$~\msyr~\kms, for the outflows driven by W51e2e, W51north, W51e8, respectively.
Using the $^{12}$CO $J=3-2$ line,  
\citet{Shi2010b} found a larger outflow mass of 1.3~\ms\ and a lower momentum rate of 0.04~\msyr~\kms\ for W51e2 (both estimates use only the blueshifted lobe). 
Their larger mass is not surprising, since $^{12}$CO  probes swept-up ambient gas on much larger scales ($\gtrsim 2$\arcsec\ or $\gtrsim 10000$~au vs. $\lesssim$ 2600~au), while a lower momentum rate is expected for a much older fossil outflow (with an age of about 1600 years vs. 115 years of the SiO outflow).

At this point we can convert the computed momentum rates of the molecular outflows into ejection rates of the primary winds/jets. This requires the knowledge of their speed, which however we do not measure. 
 Jet speeds measured in low-mass outflows via spectroscopy of atomic lines in
the optical and near-infrared wavelengths are typically a few hundred kilometers per second  \citep[e.g.,][]{Frank2014}. 
For massive outflows, there are fewer measurements of jet velocities, but a handful of proper-motion studies of the radio continuum jets provide speeds of about 500 \kms\ 
\citep[e.g., HH 80/81 -][]{Marti1995}. 
Using the total momentum rates computed above, we derive the following mass ejection  rates for the primary jet: \\
$\dot{m_j} = \left[ \frac{v_j}{500~(km~s^{-1})} \right] \times \left[ \frac{\cos{45^{\circ}}}{\cos{(i)}}\right] \times$ \  $8.6 \ee{-4}$ \msyr\   in W51e2e, 

~~~~~~~~~~~~~~~~~~~~~~~~~~~~~~~~~~~~~~~~~~~$4.8\ee{-4}$ \msyr\ in W51north, 

~~~~~~~~~~~~~~~~~~~~~~~~~~~~~~~~~~~~~~~~~~~$1.4\ee{-4}$ \msyr\ in W51e8, 

\noindent where  $i$ is expressed in units of \dg\ and $v_j$ is the jet speed in \kms. 
We emphasize that the estimate for W51e8 is less reliable  
owing to its high apparent inclination (we detect red- and blue-shifted emission on both sides of the central protostar, indicating $i<45\deg$). 

\input{table_outflows}

\section{Discussion}
\label{sect:disc} 

The analysis of the ALMA long-baselines images presented in the previous section yields four main findings:

\begin{enumerate}[i.]

\item
There are warm ($\gtrsim 50-150$~K) 
dusty filaments converging onto compact cores (Figs.~\ref{fig:rgb_overlays},  \ref{fig:dust_tb}, and \ref{fig:dust_tb_e2}) that we suggest are   accretion flows  (\S~\ref{res:acc_fil}). 

\item
We find no hints of rotation towards these cores (Fig.~\ref{fig:mom1}), suggesting that they do not host large  disks ($R_{max}<$75, $<$350, and $<$500 au for e8, north, and e2e respectively) (\S~\ref{res:disk_size}).

\item
The central sources in the cores drive young ($\sim$100~years), compact ($\lesssim$2000~au), fast ($\sim$100~\kms), powerful ($10^{-3}-10^{-4}$ \msyr), collimated SiO outflows (Fig.~\ref{fig:outflow_e2e} and Fig.~\ref{fig:outflow_n}), indicating that the driving protostars are vigorously accreting (\S~\ref{disc:accretion_rate}). 

\item
The outflow from W51north has a different axis (projected onto the plane of the sky) as a function of distance from the protostar (Fig.~\ref{fig:outflow_n}),  hinting at a change of direction over time (\S~\ref{disc:outflow_n}). 

\end{enumerate}

In the following, we  discuss these findings in detail.

\subsection{Dusty streamers as accretion channels}
\label{res:acc_fil}

The dust emission does not show a simple flattened structure at the center of the outflows, as one would expect if there is a large accretion disk \citep[c.f.][]{Ginsburg2018}, but instead the emission is resolved into multiple "streamers" or "dust lanes'' that converge onto the central cores. 
These dust lanes  appear elongated in several directions, both perpendicular and parallel, relative to the SiO outflows in all three sources.
They extend typically  across 0\pas1-0\pas4 ($\sim$500-2000 au) from the compact sources,  have relatively high 
surface brightness, indicating high intrinsic temperatures ($\gtrsim 50-150$~K; see  Figs.~\ref{fig:dust_tb} and \ref{fig:dust_tb_e2}), and carry a significant amount of mass (several solar masses; see \S~\ref{cont:mass}). 

\paragraph{Lack of kinematic signatures of accretion.}
Having identified these dusty streamers, we attempted to determine their kinematics to assess their nature, in particular to distinguish between material that is plunging into the cores (infall) or being flown out (outflow). 
Unfortunately, we do not identify any clear kinematic signatures toward them (\S~\ref{kinematics} and Figure~\ref{fig:mom1}). 
The only exception is W51e2e (Fig.~\ref{fig:mom1}, left panel), which shows red-shifted velocities along the  NW dust lane. Since these lines are seen in absorption  against the dust continuum, one would be prompted to  conclude that they probe dense gas infalling  toward the compact core.    
However, the redshifted velocities are observed  away from   the dust continuum peak and only along the redshifted lobe of the SiO outflow, which is pointing {\it away from us},  suggesting that this material may actually be entrained by the outflow. 
In this scenario, the lower opacity expected along the outflow cavity may explain why dense gas tracers like CH$_3$CN display kinematics only along that specific line-of-sight. 

\paragraph{Are the dust lanes outflow rather than accretion flows?}
The presence of CH$_3$CN along the outflow  raises the question whether the dust lanes  are actually probing warm dust heated by the outflows (e.g. by shocks), rather than accretion filaments. 
There are however at least three  counter-arguments to this scenario. 
Firstly, 
in W51e2e, there are at least four dust lanes with  similar sizes   and brightness, which appear elongated in several random directions relative to the SiO outflows (especially when considering we see a 2D projection from a 3D structure), therefore it is unlikely that they all identify dust in the outflow cavities. 
Similar arguments apply to  W51north, where there are three different lanes with no common point of symmetry.  
Secondly, 
if the outflow were responsible for the observed dust emission, the lack of bright emission toward the blueshifted lobe in the  SE from W51e2e would imply a much lower density of material in that direction, yet the blue lobe truncates sharply at a distance of about 2000~au at a point where no hot dust emission is observed.  The hot dust emission is therefore almost certainly driven by (1) the presence of (much) more gas+dust toward the  NW than the  SE and (2) proximity of that gas+dust to the central star.  Even if the outflow is interacting with this material, it is clear that there is a major asymmetry in the total mass.
 Thirdly, we suspect that 
heated dust in an outflow cavity would be too weak to be seen at the distance of W51, while the measured high surface brightness  along the filaments indicates a high intrinsic temperature. \\

Since these features are not part of an outflow, we suggest that they are  accretion filaments feeding those central sources.
Indirect evidence for infall comes from the fact that these dust lanes have a symmetric structure around protostellar sources with very high accretion rates (see \S~\ref{disc:accretion_rate}).
A second indirect evidence comes from a previously published magnetic field study with ALMA, as we detail in the next paragraph.

\paragraph{ Magnetic fields and filamentary accretion. }
The magnetic field and gravity are the two main forces that influence the dynamics of dense cores where star formation takes place. 
We compare our observations with the polarization maps of \citet{Koch2018} who conducted
a multi-scale analysis of magnetic fields in W51. The plane-of-sky component of magnetic fields inferred from the polarized dust emission exhibits a different morphology at angular resolutions of $2''$, $0''.7$ and $0''.26$ for  W51e2e, W51e8, and W51north, respectively (see Figure~2 in \citealt{Koch2018}).
At a resolution of $2''$, the magnetic field in W51e2e displays a nearly uniform distribution in the east-west direction.  The higher resolution images reveal a {\it radial} distribution of the field centered at W51e2e. The magnetic field orientations align well with the major axes of the streamers presented in Fig.~\ref{fig:rgb_overlays}.
The alignment of the dust filaments with the  radially distributed magnetic fields within the W51e2e core 
supports a scenario where the core collapse  leads to accretion flows along the filaments that drag the magnetic fields, rending a radial distribution of the field lines.
The dust emission toward the W51e8 and W51north cores are less filamentary as compared to the W51e2e core. Nevertheless, the magnetic field maps in these two cores also exhibit radial distributions 
that are suggestive of the field being dragged by the mass flow. 
This scenario has been proposed to explain magnetic field properties in dust polarization studies of high- and low-mass star formation \citep[e.g.][see also \S~\ref{implications}]{Zhang2014,HullZhang2019}. 
Future observations of kinematics in the W51 cores are needed to confirm the accretion flows in these regions.

\subsection{Upper limits to protostellar disk sizes}
\label{res:disk_size}

The compact cores at the center of the dusty streamers (i.e. towards the dust continuum peaks) is  where one would expect to see kinematic signatures of rotation (e.g. velocity gradients) if large accretion disks were present.
Although we cannot directly estimate their actual size, we can measure upper limits on the potentially disk-containing regions  based on the observed extent of the optically-thick dust emission (\S~\ref{cont:size}). This yields diameters of 
150 au for W51e8, 700 au for W51north, and 1000~au for  W51e2e, respectively (indicated with black arrows in Fig.~\ref{fig:rgb_overlays}). 

The putative compact disks are   surrounded by a morphologically flattened envelope, of size of order of 1000-2000~au, where the filaments appear to converge. 
\citet{Zapata2009,Zapata2010} used the SMA to reveal a large rotating structure on scales $\gtrsim 10000$~au. 
Although the maximum recovery scale in our study is only 0\pas4, which prevents us from recovering the large scales probed with the SMA, no such large-scale rotation was identified in the cycle 2 study with a maximum recovery scale of 9\arcsec\ \citep[][Fig 31 and 32]{Ginsburg2017}.  

\subsection{Outflow and accretion rate}
\label{disc:accretion_rate}
 Despite the lack of (large) disks, the presence of fast ($\sim \pm$100~\kms), collimated  bipolar outflows  in SiO emission   provides a clear indication of ongoing accretion onto the three HMYSOs. 
Lacking a clear signature of accretion, we cannot {\it directly} estimate the mass accretion rate. 
In jet-protostar systems, however, ejection rates are expected to correlate with accretion/infall rates.
Therefore we can use  the ejection rates calculated in \S~\ref{analysis:momentum_rate} as a proxy for  the infall/accretion rates. 
Here, we use the term ``infall rate" to describe gas in the core {\it infalling} onto the disk/protostar system, and the term ``accretion rate" to describe gas {\it accreting} onto the protostar itself. 
 Therefore we can define: 
 $\dot{M}_{infall} = \dot{M}_{ejection} + \dot{M}_{accretion}$ (assuming that the fraction of the infalling gas that is not accreted is actually re-ejected via protostellar winds/jets -- e.g., \citealt{Frank2014}).

We define $f_j = \dot{M}_{ejection}/\dot{M}_{accretion}$ as the fraction of accreting gas that is launched in the jet. 
Therefore  $\dot{M}_{infall} = (1+f_j)/f_j \times \dot{M}_{ejection}$. 
This fraction is observationally uncertain but models predict $f_j = 0.2-0.5$ \citep[e.g.,][]{OffnerArce2014,Kuiper2016}. 
Adopting these values implies $\dot{M}_{infall} \sim 3-6 \ \dot{M}_{ejection}$ and  $\dot{M}_{accretion} \sim 2-5 \  \dot{M}_{ejection}$, respectively.
Using these assumed fractions, we infer infall rates of  $2.6-5.2\ee{-3}$ and $1.4-2.9\ee{-3}$~\msyr\, and accretion rates of $1.7-4.3\ee{-3}$ and $1.0-2.4\ee{-3}$~\msyr, onto the W51e2e and W51north protostars, respectively\footnote{The outflow in W51e8 appears to be nearly in the plane of the sky, making estimates of the infall/accretion rates less reliable.}.  

\subsection{Outflow from W51n: a single outflow changing direction on the sky}
\label{disc:outflow_n}

\begin{figure*}
\includegraphics[width=0.95\textwidth]{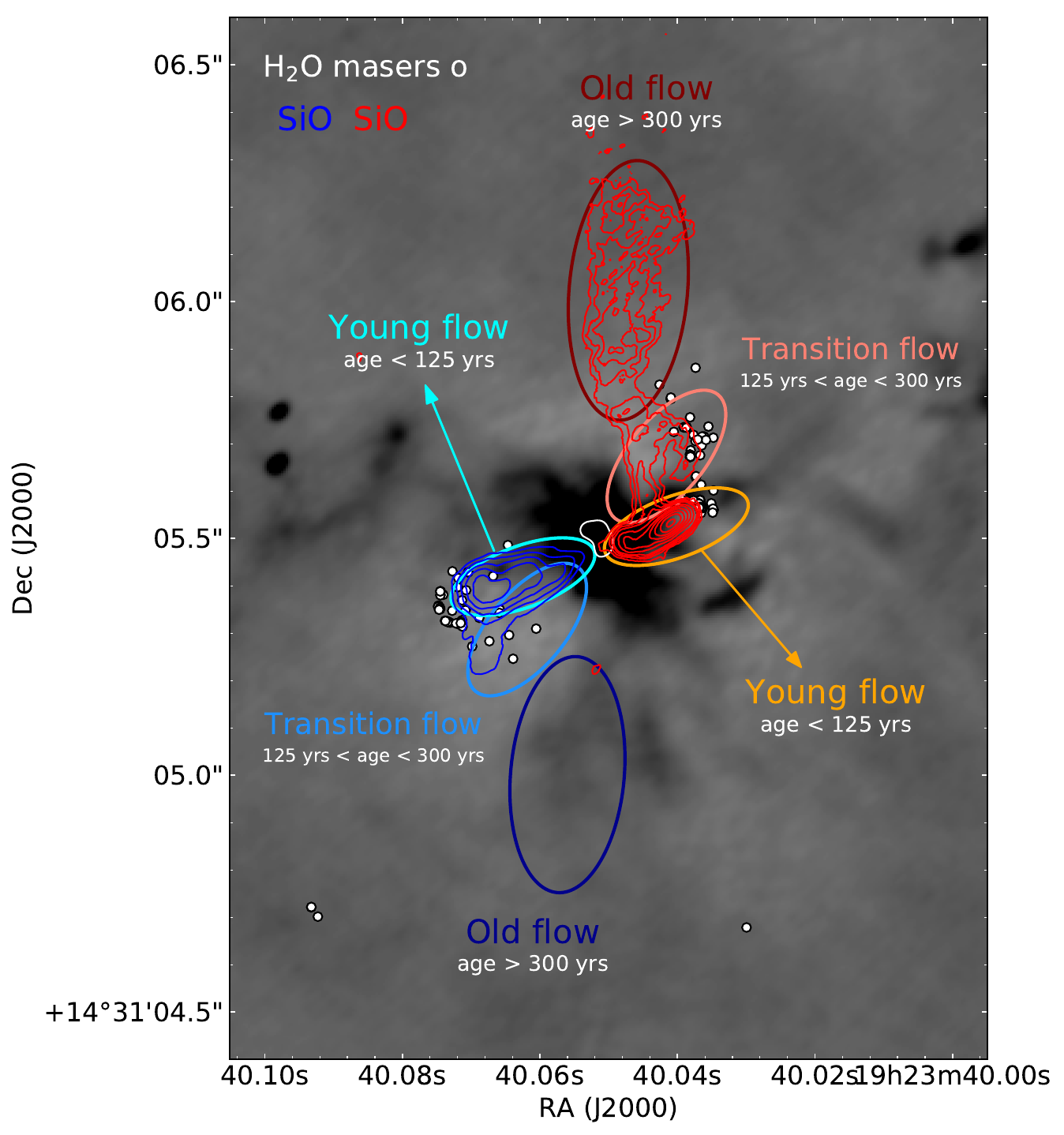}
\caption{Sketch of the history of the outflow from W51north.
The  H$_2$O masers positions (white circles) measured with the VLBA \citep{Imai2002} are overplotted  onto the total intensity of the redshifted (red contours) and blueshifted (blue contours) emission of  the SiO J = 5-4 line and the $\lambda$1.3~mm continuum emission (greyscale image). 
The color ellipses identify  three outflow components tracking its history: the `old' outflow ($t> 300$ yr) 
 shows a fossil redshifted lobe (and a now invisible  blueshifted lobe) along north-south (indicated by the dark red and dark blue ellipses);  the `young', bright outflow ($t<125$ yr) is the present outflow with an axis along  SE- NW (indicated by the  orange and cyan ellipses); 
  the `transition' flow (with a total duration of about 175 years) represents material ejected while  the outflow changes directions from PA --13\dg\ to --70\dg\ (identified by light red and light blue ellipses). 
The lack of the `old' blueshifted counterpart to the `old' redshifted lobe may be due either to some sort of localized blockage (so the outflow never blew out) or to the lack of material to be entrained (e.g., the outflow breaks out of the cloud). It is worth noting that even in the lower-resolution maps of the CO 2-1 emission, the blueshifted lobe is barely detected beyond about 0\pas30 or 1600~au  \citep{Ginsburg2017}.  
 }
\label{fig:outflow_n_sketch}
\end{figure*}

In \S~\ref{res:outflows} we show that the W51north outflow appears to have changed direction substantially, as measured with a change of P.A. (projected onto the plane of the sky) from $\sim-$10\dg~to $\sim-$70\dg, from scales $\gtrsim$2000--20000~au down to the smallest scales $\lesssim$200--1000~au.    
Here we argue that this `multi-component' morphological structure  is best explained as a single  protostellar outflow that has changed  direction  over a short period of time (but see possible alternative scenarios in \S~\ref{outflow_n_alternatives}).
In this scenario, 
the prominent and compact  SE-NW (P.A.$\sim-$70\dg) faster component represents the  recently-launched outflow, while the faint and diffuse  north-south (P.A.$\sim-$10\dg) slower component represents a fossil outflow. 
In particular, 
assuming the outflow was driven at a constant velocity $v_{outflow}=30$ \kms\ (the average velocity of the bulk of the SiO emission), 
we infer the following history of the outflow:
\begin{itemize}
    \item {\it "Old"  flow}: From about 22,000 au to 1900 au as traced by CO (and from about 4300 as traced by SiO) the outflow was consistently pointed at P.A. $-13\deg$  (3500 to 300 years ago).
    \item {\it "Transition" flow}: From 1900 to 800 au, the outflow changes directions from P.A. $-13\deg$ to $-70\deg$ (300 to 125 years ago, duration 175 years).
    \item {\it "Young"  flow}: From 800 au to the resolution limit, the outflow has been consistently pointed along $-70\deg$  ($\sim 125$ years ago to now). 
    
    \vspace{0.1cm}
    This outflow history is sketched in Fig.~\ref{fig:outflow_n_sketch} for the SiO portion of the outflow. 
    (See also Appendix~\ref{app:outflow_age} for  details on the dynamical age estimates.)
\end{itemize}

The outflow morphology fits this scenario: there is emission detected in both the red and blue lobes at intermediate position angles ($-10^\circ > \rm{P.A.} > -70^\circ$).
This emission is not as cleanly symmetric as the more linear ``old'' and ``young'' outflows, which may be caused either by asymmetries in the surrounding medium or by changing speed of the outflow launch.
A point in favor of this scenario is the lack of detected outflow emission in either SiO or CO \citep{Ginsburg2017}  beyond 1300~au along the current  SE-NW outflow axis.
If the W51north outflow has not changed directions in the last $\sim100$ years, the lack of observed outflow tracers at greater distances along  SE-NW implies that accretion onto this source only began in the last 100 years.
Given its high luminosity ($L\gtrsim10^4$ L$_\odot$), such an age is implausible.  

The dramatic change of direction  of the W51north outflow  suggests a sudden, major event that redirected the driving jet over the course of just 150-200 years. 
This in turn indicates a substantial change in the orientation of the outflow-launching region, i.e., of the accretion disk, over that same timescale.
The large change in angle ($\gtrsim 60$\dg) suggests that small perturbations or instabilities in the disk are inadequate.
Given the presence of a substantial reservoir of surrounding material distributed asymmetrically around the source, we argue that accretion onto the disk is a likely cause for the change of outflow direction.
Under the assumption that the present (young) outflow is the result of a single major accretion event and that the  infall rate remained constant over the transition period (175 years), from the infall rate estimated in \S~\ref{disc:accretion_rate}, 
we infer that  a mass of about 0.25--0.5~\ms\ was dumped onto the protostar/disk system in such accretion event.  

Is this mass enough to flip the disk?
We assume that the disk has a radius of 350 au and  a mass of 1~\ms\ (i.e.,  a quarter of the compact core mass is in the disk; see Table~\ref{tab:src_masses}), and is in Keplerian rotation about a 20~\ms\ central star.
The total angular momentum in such a disk would be $L_{\rm disk}\sim 3 \times 10^{54}$ g cm$^2$ s$^{-1}$ (see calculation in Appendix~\ref{app:diskL}).
If a 0.25--0.5~\ms\ condensation impacts the disk at a radius of 350 au with a velocity of 10~\kms\ (this corresponds to the infall velocity at 350 au toward a 20~\ms\ star),  the angular momentum of the impactor is $L_{\rm dump} \sim 2.6-5.3 \times 10^{54}\ \sin  \alpha$ g cm$^2$ s$^{-1}$, where $\alpha$ is the angle between the impactor trajectory and the disk plane.
For $\alpha = 90$\dg, 
the disk angular momentum vector is reoriented by $\theta = \tan^{-1}  ( L_{\rm dump}/L_{\rm disk} ) = 41-61$\dg\ (for a smaller $\alpha$ the flip would be  smaller by a factor $ \sin \alpha$).
We explicitly notice that $\theta$ is independent from the protostellar mass, since both  $L_{\rm dump}$ and $L_{\rm disk} \propto \sqrt{M_*}$.  
We conclude that the accretion-driven disk reorientation model is plausible.

%
\subsubsection{Alternative scenarios}
\label{outflow_n_alternatives}

We now consider alternative scenarios to explain the peculiar structure of the  SiO outflow from W51north: 

\begin{enumerate}
\item The  outflow hits an ``obstacle" in the surrounding dense clump and gets deflected in another direction (one jet-driving YSO).
\item  There are two independent outflows driven by a binary with a separation $<$2000 au (two  jet-driving YSOs).
\item  There are three independent outflows driven by three different protostars within 2000 au (three  jet-driving YSOs).
\item There is a single precessing outflow in a  protostellar  binary  (one  jet-driving YSO and one companion).
\item A close stellar encounter (a fly-by) changes the 3D orientation of the disk/jet system   (one  jet-driving YSO and one intruder). 
\end{enumerate}

\paragraph{Deflection by an obstacle. }
This scenario requires the presence of two obstacles located at approximately the same distance from the central source along the axis of the outflow, and these obstacles must be angled such that the redshifted flow is deflected to the north and the blue flow to the south.  Such a contrived scenario is implausible, since it requires a unique and unlikely set of conditions with no known analogs in the observational or theoretical literature.

\paragraph{Two independent outflows from a binary. }
Two  protostars in a binary, with a separation $<$2000 au  and surrounded by two disks, would drive two independent outflows, seen on different spatial scales in CO and SiO line emission, respectively. Similar cases are known in the literature (e.g., the L1551 binary in the Taurus SFR - \citealt{Rodriguez2003}). In this case we can rule out simultaneous independent outflows because we do not see high-velocity material close to the star in the north-south direction, implying that the north-south outflow is no longer there.  The apparent continuity between the smaller NW/SE flow and the larger north-south flow is evidence against this hypothesis, since there is no reason these independent flows would be connected on large scales.

\paragraph{Three independent outflows driven by three different YSOs. }
The powerful NE-SW  outflow, the diffuse  northern flow, and the faint southern flow could be driven by three independent YSOs. 
This scenario requires at least two undetected YSOs to be symmetrically offset from the central source and be precisely at the locations of the intersections between the SiO and CO outflows. 
Each of these YSOs would have to drive a single-lobe outflow\footnote{Single-lobe outflows have been observed, and can be explained with inclination effects and/or with different densities of ambient material around the source.}
in the blueshifted (south) and redshifted (north) direction.  
These single-lobe flows must also have velocities consistent with the brighter SiO outflow from the central source (redshifted for the northern YSO and blueshifted in the southern YSO). 
The detection of the W51north, e2e, and e8 outflows in SiO, but nondetection of other SiO outflows in the field that contains dozens of candidate YSOs confirms that such high-brightness SiO outflows are very rare, so that even if there were two perfectly positioned YSOs, they are each unlikely to drive an SiO outflow.
Combined, the various restrictions imposed by this scenatio make it highly improbable.

\paragraph{Precessing outflow in a tight binary. }
 Theoretical models  suggest that gravitational instabilities during the core collapse would cause the disk to fragment, resulting in massive binary or multiple systems, depending on the initial mass of the cloud \citep[e.g.][]{Krumholz2009}. 
 Tidal interactions between the disk associated with the primary and a companion star in a noncoplanar orbit would naturally lead to precession of  the  disk/jet system associated with the primary. 
   Precessing outflows have been observed  in massive binaries. 
   Two well-known examples are the two archetypal high-mass (B-type) YSO binaries IRAS~20126+4104 and Cepheus-A. 
In both objects there is evidence of steady precession over several thousand years possibly driven by a binary orbit that affects the disk \citep[e.g.,][]{Cunningham2009}.
Qualitatively, the existence of these precessing outflows confirms that a single-source outflow changing direction over time is possible.
 Although we cannot exclude  the existence of a binary system at the center of W51north, the spatial and kinematic structure of the SiO/CO outflow is inconsistent with the expected jet precession.

We can use IRAS~20126+4104   as a benchmark to compare its observational properties  to W51north. 
 There are some significant structural and kinematical differences between the two sources.
  First, W51north does not show an S-shaped morphology on large scales \citep[][]{Shepherd2000,Ces05} nor wiggling on small scales \citep{Caratti2008}, as observed in IRAS~20126+4104, but rather displays two different outflow axes, one along a NW-SE direction (within 1600 au  from the protostar), and one along the north-south direction (up to scales of 0.5 pc). This is inconsistent with a regular jet precession due to a binary and/or a multiple, but it is consistent with a sudden change of disk/outflow direction from a single YSO.   
  Secondly, while in IRAS 20126+4104 the large-scale north-south outflow   appears poorly collimated, as a  consequence of the severe precession of the jet, in W51north the two outflow components appear bipolar and collimated on both (small and large) scales.   
Thirdly, in IRAS~20126+4104, 
the velocity of the lobes reverses so that, the blue-shifted gas is located to the SE at low velocities, whereas it appears to the NW at high velocities. 
 In W51north we do not see such a velocity reversal.

  In summary, unlike IRAS~20126+4104  and Cepheus A, W51north exhibits a single shift in direction that occurred on a shorter timescale (when compared with the thousands or tens of thousands years of their precessing jet), suggesting that a steady binary interaction is not the driver. \\

\paragraph{A stellar intruder hits the disk.} 
Since W51 is forming a dense protocluster \citep{Ginsburg2016,Ginsburg2017}, it is likely that
W51north formed in the presence of lower-mass  protostars (see also discussion in the last paragraph of \S~\ref{implications}).
If a close passage between a lower-mass  protostar and W51north has occurred about 150-200 years ago then it is possible that the disk has been  reoriented. 
Although we cannot  rule out this scenario, 
 the most probable effect of a close passage would be disk disruption \citep[e.g.,][]{MoeckelGoddi2012}, explosion-like events \citep[e.g.,][]{Zapata2009, Bally2020} and/or stellar ejections (if more than two stars are involved; e.g., \citealt{Goddi2011a,Rodriguez2020}). 
 The nicely collimated bipolar outflow on scales $<$1600~au, hinting at an an ordered disk, argues against this scenario. 
Even assuming that the disk would survive, the low collisional cross-section between the disk and a star (whose effect is limited to the star's orbital radius, as opposed to a gas inflow) would require several orbits, and therefore several close passages, to completely transfer its angular momentum to the disk/jet system. Given the short dynamical time of the outflow turn ($<200$ years), this scenario is unlikely. 
\\

We conclude that the scenario where a single YSO changes its disk/jet orientation as a consequence of substantial accretion event does not suffer from any of the improbability of the  alternative scenarios considered here, and it is consistent with the observed morphology in several unique ways,  therefore providing the best explanation for the W51north outflow.

\section{General implications  on the high-mass star formation process}
\label{implications}

The observational findings presented in this ALMA study  support a scenario where accretion filaments are feeding  compact (unsteady) disks, which drive collimated outflows that can change orientation over time.  
Such compact disks should also change their orientation, which implies loss of angular momentum support, probably due to varying angular momentum of the accreting material. The latter suggests episodic accretion and periods of (very) high accretion.
Such  a scenario has some interesting  implications  on our understanding of the high-mass star formation process, which we detail below. 

\paragraph{ Multi-directional  unsteady accretion. }

Modern 3D (M)HD simulations predict that accretion onto protostellar cores proceeds highly asymmetrically  along filaments on scales greater than about 100 au \citep{Cunningham2011,Commercon2011,Myers2013,Seifried2015,Rosen2016,Klassen2016,Rosen2019,Rosen2020}. 

  For instance,  \citet{Seifried2015} performed magneto-hydrodynamic simulations of turbulent and magnetized collapsing cores and showed that accretion of mass and angular momentum is highly anisotropic and proceeds through a few narrow and strongly pronounced channels on scales of order of  1000 au. These channels are reminiscent of the  filamentary structure revealed by the dust continuum emission in our  ALMA maps (e.g., compare their Figures 1 and 6 with the left panel of Figure 1).
\citet{Rosen2016} added radiation feedback and their 3D radiation-hydrodynamic  simulations  show that most of the mass is supplied  to the accreting star  via gravitational instabilities in dense filaments and only when  an extended disk is formed, the majority of mass delivered to the massive star is due to disk accretion. 
Similarly, \citet{Klassen2016} performed radiation-hydrodynamic simulations of collapsing protostellar cores with different initial masses (up to 200 \ms), 
and noticed that after a small protostellar disk starts forming (after 15 kyr), it becomes asymmetric and gravitationally unstable (after 25 kyr) and develops spiral density waves which act as accretion channels (with size of order 1000 au).
Their simulations show steady inward gas motion across virtually all angles (besides accretion in the disk plane). They also find that the protostar launches  powerful outflows with velocities exceeding 100 \kms.

Our observations have achieved the angular resolution and sensitivity needed to  identify these {\it multi-directional} accretion  structures on scales 100-2000 au and provide an experimental confirmation to the predictions of modern  (M)HD simulations.

\paragraph{ Episodic accretion and  transient  disk/outflow systems. } 

A natural consequence of the multi-directional nature of the accretion flows is that the dominant angular momentum direction does not remain constant over the course of the mass assembly process \citep{Smith2011,Rosen2020}. This in turn implies the occurrence of unsteady, episodic accretion events onto the disk and subsequently onto the protostar itself \citep{Meyer2019}. 
If the parent core feeding the central disk has a distribution of angular momentum vectors at different radii from the protostar, the central disk will accrete gas with time-dependent directions of the mean gas angular momentum vector. 
Irregular accretion may result in sudden changes of the orientation of the disk/jet when a large amount of material is suddenly accreted onto the disk.
Since the disks reside at the center of larger scale filaments, 
the buildup of a steady protostellar disk can occur only during periods when the accreting filaments do not significantly change their orientation relative to the disk.
In W51north, a dramatic structural change in the orientation of the accretion channels and/or a large accretion event may have resulted in a sudden change of orientation of the accretion disk.

In the low-mass regime, evidence of episodic accretion is observed in  FU Ori objects \citep[e.g.,][]{HartmannKenyon1996,audard2014}, which exhibit sudden  increases in the accretion rates (and therefore luminosities) of a few orders of magnitude that last from a few tens of years to a few centuries.
Recently, a few outbursts in the high-mass regime have also been reported  in the literature.  These events are marked by substantial increases in maser \citep{Moscadelli2017} and dust emission both at mm \citep{Hunter2017} and/or IR \citep{Caratti2017} wavelengths. 

The rapid and substantial change in the disk orientation observed in W51north   
 provides an indirect observational evidence of a significant recent accretion burst in the growth of a massive protostar.

\paragraph{ Magnetic fields and outflow orientation. } 

In alternative, the outflow from W51north would not be a classic magnetocentrifugal wind  \citep{BlandfordPayne1982} but a "spiral" outflow \citep{Matsumoto2017}. 
 In the latter, the outflow is not necessarily aligned with the rotational axis of the disk.
In fact, since  circumstellar disks are aligned according to their angular momentum,
while  outflows are elongated in the direction parallel to the local magnetic field, 
if the latter had different structures on different scales, the outflow could have different orientations on these different scales. 
This mechanism could readily produce a disk that is misaligned with the outflow 
and/or cause the outflow closer to the protostar to be misaligned with the components at larger scales. 
Misalignment between  disks, outflows, envelopes, and magnetic fields  have been observed in low-mass and high-mass star forming regions \citep[e.g.,][]{Hull2013,HullZhang2019}.

\paragraph{Magnetic fields and filamentary accretion.}  
In a survey of 14 high-mass star forming regions with the SMA, \citet{Zhang2014} found that magnetic fields at 0.1~pc cores tend to be aligned with the field in the pc-scale parental clump. Based on this statistical study, they concluded that the magnetic field is dynamically important during the fragmentation of the clump and the formation of dense cores. In addition, they found that the major axis of protostellar outflows do not appear to correlate with the magnetic field orientation in dense cores, which suggests that gravity and the angular momentum may dominate the magnetic field from 0.1~pc core scales to the 100~au disk scale 
\citep[see also][for a review]{HullZhang2019}.
The alignment of the dusty streamers with the radial magnetic field  orientations   within  the  W51e2e  core (and to some extent in W51e8 and W51north cores) is  consistent with  the  findings  in  \citet{Zhang2014} and suggests a scenario where accretion flows along the filaments  drag the magnetic fields, as suggested in \S~\ref{res:acc_fil}.  

\paragraph{  Compact   accretion disks.} 

Even with an unprecedented angular resolution of 20 mas, we do not identify disks in our protostellar sources, mainly because of  high optical depth of the dust continuum emission. 
Nevertheless, we believe disks are present based on collimated SiO outflows, and we estimate upper limits on the disk radii of $<$75~au for W51e8, 350~au for W51north, and 500~au for  W51e2e, respectively. 
These upper limits are consistent with  predictions from recent (M)HD simulations. 
In particular, simulations producing asymmetric accretion flows suggest that the disks should only form on scales less than about 100~au  \citep{Commercon2011,Myers2013,Seifried2012,Seifried2013,Seifried2015}, whereas 
  simulations that have more ordered initial conditions and do not produce asymmetric flows \citep{Krumholz2009,Kuiper2011,Klassen2016} often produce much larger ($\sim500$ au) disks.

We can now compare the disk size upper limits derived in this study with the existing measurements of disks in other HMYSOs. 
In Table~\ref{tab:HMdisks} we list all known Keplerian disk candidates around HMYSOs based on ALMA observations todate.  
The majority of disks have radii in the range 300--1000~au.
Most of these disks would be resolved in this ALMA long-baseline data set at $d_{W51}=5.4$~kpc; only the smallest few would be missed.

{\it So why are the disks larger in other HMYSOs? }

 In an attempt to address this question, in the following we will explore two additional star formation properties: evolution and environment.

\paragraph{ Evolutionary stage and disk detection rate. }  

The lack of large disks in W51, as well as the low disk detection rate around HMYSOs, could be related to their evolutionary stage. 
\citet{Cesaroni2017} explored the properties of seven HMYSOs hosting rotation disk candidates using ALMA at 0\pas2 resolution. In particular, they plotted the luminosity-to-gas mass ratio (an evolutionary stage indicator) as a function of the distance for all seven HMYSOs, and concluded that the disk detection rate could be sensitive to protostellar evolution. 
In particular, 
in young, deeply embedded sources, the evidence for (Keplerian) disks could be weak because of confusion with the surrounding infalling envelope or toroid \citep[e.g.,  G351.77-0.54 --][]{Beuther2017}, while in the most evolved sources the molecular component of the disk could have already been greatly reduced or completely dispersed, resulting either in very small disks with $ r <50-100$~au (e.g., Orion Source I - \citealt{Ginsburg2018} - and  G17.64+0.16 - \citealt{Maud2019}) or no disk at all. Only in those objects that are at an intermediate stage of the evolution, the molecular gas emission could reveal the presence of an underlying disk. 
This evolutionary trend is supported by the numerical simulations cited above, that predict that Keplerian disks
are constantly being fed material from large-scale infalling filaments and 
grow with time {\it from the inside out}, from tens of au up to $\sim$1000~au \citep[e.g.,][]{Kuiper2011,Rosen2016,Klassen2016}.

\paragraph{Evolutionary stage and ionization rate. } 
One typical indicator of evolution in HMSF is provided by radio continuum emission produced by photoionization induced by Lyman photons. 
When the HMYSOs reach the ZAMS, they produce copious amount of ultraviolet radiation and start ionising their surroundings, forming compact \hii\,regions. 
HMYSOs can also exhibit radio continuum via shock ionization in jets/outflows \citep[e.g.][]{Moscadelli2016,Sanna2018}. 
The  HMYSOs listed in Table~\ref{tab:HMdisks} are all known to exhibit radio continuum emission, either in the form of an ionized wind/jet or a compact \hii\,region, indicating that they are relatively evolved\footnote{G351.77-0.54 is an exception since, to the best of our knowledge, it does not produce radio continuum emission.}. 
The non-detection of radio continuum emission, and in particular the lack of an observable  \hii\,region, towards the three HMYSOs studied here suggests that 
they  may still be in the pre-main-sequence and can therefore be considered the high-mass counterparts to the ``Class 0" stage in solar-like stars. 
 This could explain why we do not see large disks. 
 
\paragraph{Evolutionary stage and high accretion rate. } 
In alternative, the non-detection of radio continuum emission could be a consequence of very high-density accretion flows. 
 In fact, detailed theoretical studies  suggest that the evolution of a high mass protostar depends strongly on the accretion rate onto the protostellar surface and the exact accretion geometry \citep{Keto2003,Hosokawa2010}. 
 For example, a high accretion rate ($\gtrsim 10^3$~\msyr),  
as inferred for  the three HMYSOs in W51 (and in general expected for high-mass star formation; e.g., \citealt{ZinneckerYorke2007}), 
could result in 
a quenched or trapped \hii\,region \citep{Keto2003,Keto2006}.
This scenario requires however spherical symmetry and it is therefore disfavored by our data which show highly asymmetric accretion flows. %
In alternative, high accretion rates 
can change the properties of the underlying star, bloating it and reducing its effective photospheric temperature \citep{Hosokawa2009,Smith2012,Hosokawa2016,Tanaka2016}.
This significantly delays stellar ignition (up to 20~\ms), changing the history of the energy feedback (e.g., bolometric luminosity, total amount of Lyman photons, etc.).  
Therefore, the lack of an observable \hii\,region could actually be explained with high accretion rates in a bloated HMYSO. 
We explicitly note that this mechanism has been  proposed to explain  the low luminosity ($\sim 10^4$~\ls) estimated for G11.92-0.61 MM1, which, with a mass  of $34\pm5$~\ms\ \citep{Ilee2018}, is the most massive O-type Keplerian disk candidate known to date (see Table~\ref{tab:HMdisks}).

\paragraph{ Protostellar disk evolution and protocluster environment. } 

A final  question worth addressing is  whether the two properties of the disks we have unveiled  (small size and changing orientation) are intrinsic to the HMSF evolution or due to the dense cluster environment. 
On pc-scales, W51 contains ten-fold more mass than other nearby regions \citep[e.g.,][]{Ginsburg2016,Ginsburg2017}, and, assuming a standard star formation conversion rate, we would expect to form ten times more stars relative to a typical 1000 \ms/pc scale clump \citep{Ginsburg2013}. 
So the three HMYSOs studied here are likely forming in the presence of lower mass condensations and/or lower mass stars that could have been interacting with them. 
Simulations of such clustered environments find that boosting of the accretion rate can be induced with protostellar encounters     \citep{Pfalzner2008} or even mergers \citep{BonnellBate2005}. 
It is an open question  whether the dense stellar environment is affecting the protostellar (disk) evolution on the scales resolved in this study, i.e.  100-2000~au \citep{Rosen2016,Rosen2019}.

\section{Summary and Conclusions}
\label{sect:conclusions}

Using the ALMA longest baselines (16 km) at  1.3~mm (Band 6), we have mapped the 
 innermost regions around three deeply embedded high-mass protostars belonging to  the W51 star forming complex with very high spatial resolution ($\sim$100--200 astronomical units).  
 
 We summarize the main findings  of this study as following:
 
 \begin{enumerate}[i.]

\item
The high-angular resolution ($\sim$0\pas03 beamsize) ALMA maps of the dust continuum emission  reveal complex filamentary structures that   are $\sim$ 2000~au long and converge onto compact cores.
No clear kinematic information can be extracted towards  these filaments or the compact cores, most likely because of high optical depth in the dust continuum.  

\item
The central sources in the cores drive young ($\sim$100~years), compact ($\lesssim$2000~au), fast ($\sim$100~\kms) collimated SiO outflows.
From outflow ejection rates, we  infer accretion rates of $1.7-4.3\ee{-3}$ and $1.0-2.4\ee{-3}$~\msyr, onto the W51e2e and W51north protostars, respectively, indicating that they are vigorously accreting. 

\item     
We measure upper limits on the disk radii of $<$75~au for W51e8, 350~au for W51north, and 500~au for  W51e2e, respectively. 

\item     
The outflow from W51north displays a multi-component morphological structure.   We argue this structure is best explained as a single  protostellar outflow that has changed  direction  over a short period of time as a consequence of a substantial accretion event onto its disk. 

 \end{enumerate}

The accretion/outflow structures observed in the W51 HMYSOs appear much more complex than generally observed in  class 0 protostars \citep[but see  for instance][for examples of complex structures in Class 0 envelopes]{Tobin2012}.
  Their properties are also  different from those inferred from  Keplerian-like disks  recently reported around OB-type HMYSOs, which generally appear to be at a later evolutionary stage and are forming in less dense, clustered environments.
 
 In order to interpret the observed circumstellar structures, we propose the following scenario:

 \begin{enumerate}[a.]

\item 
The observed small-scale dusty streamers are imprints of accretion columns towards the central HMYSOs.  While there is no kinematic data to confirm this hypothesis, it is the simplest explanation consistent with both the continuum structure and the high outflow rates. 

\item 
These accretion channels inhibit the formation of large, steady disks (at least in the early stage of formation) but feed central compact disks, whose presence is inferred from the existence of massive and powerful bipolar outflows. 

\item 
These compact disks are not the classic smooth, steady disks seen around low-mass stars and in rotating core simulations, 
but instead are truncated and change orientation over time, as seen in more turbulent simulations.

\item
Outflows changing direction mark
 the occurrence of unsteady, episodic accretion events onto the disk and subsequently onto the protostar itself, leading to  accretion and luminosity bursts \citep{Meyer2017,Meyer2019}.

 \end{enumerate}

When put together, these findings contrast with a simplified vision  of an ordered disk/jet geometry and  
point to a different type of accretion that we call ``multi-directional accretion", where the accreting material has a different angular momentum vector over time.

Further high-resolution observations at lower frequencies with ALMA or in the future with the ngVLA, 
are required to elucidate the nature of the dusty streamers and identify disks in the central cores by unveiling their dynamics. 
Similar studies targeting the most massive hot cores known in the Galaxy would allow us to assess whether this multi-directional accretion mode is the standard route for the formation of the most massive stars in dense stellar environments.

\vspace{1cm}

\textbf{Acknowledgments}
This paper makes use of the following ALMA data: 2013.1.00308.S 
and 2015.1.01596.S.
ALMA is a partnership of ESO (representing its member states), NSF (USA) and
NINS (Japan), together with NRC (Canada), NSC and ASIAA (Taiwan), and KASI
(Republic of Korea), in cooperation with the Republic of Chile. The Joint ALMA
Observatory is operated by ESO, AUI/NRAO and NAOJ.
C.G. as the PI of the ALMA project 2015.1.01596.S acknowledges assistance from Allegro, the European ALMA Regional Center node in the Netherlands.
C.G. acknowledges financial support from ERC synergy grant 610058. 
 L.A.Z. acknowledges financial support from CONACyT-280775 and UNAM-PAPIIT IN110618 grants, Mexico. 
A.G. acknowledges support from the National Science Foundation under grant No. 2008101.

\bibliography{biblio}{}
\bibliographystyle{aasjournal} 

\appendix
\input{appendix}

\end{document}

%% file: commands.tex
\newcommand{\pas}{$\rlap{.}^{\prime\prime}$}
\newcommand{\dg}{$^{\circ}$}
\newcommand{\kms}{km~s$^{-1}$}
\newcommand{\cmc}{cm$^{-3}$}
\newcommand{\cmq}{cm$^{-2}$}
\newcommand{\hii}{H\,{\sc ii}}
\newcommand{\ms}{$M_{\odot}$}
\newcommand{\ls}{$L_{\odot}$}
\newcommand{\rs}{$R_{\odot}$}
\newcommand{\msyr}{$M_{\odot}$~yr$^{-1}$}
\def\kmsy{km~s$^{-1}$~yr$^{-1}$}
\def\kmo{km~s$^{-1}$~mas$^{-1}$}
\def\kmt{km~s$^{-1}$~mas$^{-2}$}
\def\vsys{V$_{\rm sys}$}
\def\jyb{Jy\,beam$^{-1}$}
\def\mjyb{mJy\,beam$^{-1}$}
\def\ms {\hbox{M$_{\odot}$}}
\def\ls {\hbox{L$_{\odot}$}} 
\def\n {$n_{\rm{H_{2}}}$}
\def\vlsr{V$_{\rm LSR}$}
\def\rsun{R$_{\odot}$}
\def\co{$^{12}$CO}
\def\wat{H$_2$O}
\def\h2{H$_2$}
\def\iras{IRAS~20126+4104}

\newcommand{\perbeam}{\ensuremath{\mathrm{beam}^{-1}}}
\newcommand{\hh}{\ensuremath{\mathrm{H}_2\xspace}}
\newcommand{\persc}{\ensuremath{\textrm{cm}^{-2}}\xspace}
\def\ee#1{\ensuremath{\times10^{#1}}}

%% file: table_src_masses.tex
\begin{deluxetable*}{lcccccccccc} 
\tablecaption{Properties of the 1.3~mm dusty sources associated with HMYSOs W51e2e, W51e8, W51north.}
\label{tab:src_masses}
\centering                        
\tablehead{
Source &~~&\multicolumn{2}{c}{Central Beam} & \multicolumn{2}{c}{Compact Core} & \multicolumn{2}{c}{Extended Structure} &Dust & & Protostellar \\
&~~& Radius  & Mass & Radius & Mass &Radius & Mass & T$_B$ Peak & & Luminosity$^{d}$  \\
&~~& [ AU ] & [ \ms ] & [ AU ] & [ \ms ] & [ AU ] & [ \ms ] & [ K ] & & [ \ls ] 
}
\startdata
W51e2e$^{a}$ &~~& 70  & 0.36 & 500  & 4--12 & 2700 & 25 & 575 && $>8.2 \times 10^3$ \\
W51e8$^{b}$ &~~& 75  & 0.36 & --  & -- & 2500 & 16 & 435 && $>7.5 \times 10^3$ \\
W51north$^{c}$ &~~& --  & -- & 350  & 4--5 & 1200 & 14--28 & 390 && $>2.8 \times 10^3$  
\enddata
\tablecomments{The sizes  and masses  are estimated for different components of the dust continuum emission. \\
(a) In W51e2e the mass estimate in the extended filamentary structure does not include the inner core.  \\ 
(b) In W51e8 there is  no core-like symmetric structure surrounding the bright central source.  \\
(c) W51north does not contain a central bright  point source.\\  
(d) These protostellar luminosities are lower limits as inferred in \S~\ref{cont:lum}. 
}
\end{deluxetable*}

%% file: table_outflows.tex
\begin{deluxetable*}{lccccccccc} 
\tablecaption{Outflow parameters derived from the SiO (J = 5--4) line emission in W51north, W51e2e, and W51e8.}
\label{tab:outflows}
\centering                        
\tablehead{
Source &  $\Delta V$ & $V_{max}$ & $R_{max}$ &  $T_{dyn}$ & I$_{\rm int}$ & M & P & $\dot{P}$  & $\dot{M}_{ejection}$  \\
& [ \kms ]  & [ \kms ] & [ AU ] & [ years ] & [ K~\kms ] & [ \ms ] & [ \ms~\kms ] & [ \msyr~\kms ] & [ \msyr ] 
}
\startdata
W51north$^{a}$ & [-36,56] &  95  & 1540 & 77 & $1.39\ee{4}$  & 0.18 & 9.2 & 0.12 & 2.4\ee{-4}\\
W51e2e$^{a}$ & [-56,53] &  105  & 2560 & 116  & $1.34\ee{4}$& 0.36 & 24.8 & 0.21 & 4.3\ee{-4}\\
W51e8$^{a}$ &[0,60]&  42  & 390 & 44  & $0.94\ee{4}$ & 0.11 & 1.5 & 0.03 & 0.7\ee{-4}\\
\enddata
\tablecomments{The parameters are estimated  from the blueshifted lobe. The total outflow mass and rate are twice the quoted values.\\
 To convert from SiO column to \hh\ column, we assume a conservatively high abundance of SiO, $X_{SiO} = N(SiO) / N(\hh) = 10^{-7}$. A lower $X_{SiO}$  would imply higher mass and momentum in the outflow. \\
The age estimates assume an inclination angle of 45\dg. For an  inclination $i$, the values need to be multiplied by a factor $\cos{(45^{\circ})}/\cos{(i)}$. 
}
\end{deluxetable*}

%% file: appendix.tex
\section{Why do not we see kinematical signatures of accretion?}
\label{app:lack_of_kin}
We have looked for expected signatures of accretion onto a high-mass, high-luminosity protostar embedded in a core that is optically thick in the continuum, in particular:
a) higher excitation lines should be redshifted compared to lower-excitation lines (warmer gas should be moving with higher velocities toward the center),
b) the spectral profiles should be moderately asymmetric (skewed toward the blueshifted component; e.g., \citealt{Goddi2016}).
The velocity field appears however to be approximately flat across the core for the three sources.
While some lines show some signs of different-velocity absorption features, these are all low-excitation lines (e.g., $^{13}$CO, H$_2$CO) that demonstrably trace the broader molecular cloud, not the core.

If there are genuinely massive, accreting cores in these objects, as demonstrated by the presence of powerful outflows, some infall \emph{must} be occurring.  
A few solutions are possible for its non-detection:
\begin{enumerate}
\item 
{\it All gas is accreting in the plane of the sky.}   
Since we see three sources, none of which have clear kinematic accretion signatures, and all of which have detectable outflow kinematic signatures, this possibility is implausible.
\item
{\it  Radiative transfer effects are hiding the kinematic signatures. }
In low-mass cores, (double-peaked) symmetric line profiles are expected in optically thin tracers because the velocity of the flow increases toward the star, and inverse P Cygni profiles in optically thick lines.  However, in low-mass cores, the continuum is almost always optically thin and has relatively low brightness temperature.
In the high-mass cores we observe, the continuum brightness temperature is very high, leading us to infer that the optical depth is similarly high.
As we probe closer to the star (where the velocities are expected to be higher), the gas temperature rises and approaches equilibrium with the dust.
If $T_{gas} = T_{dust}$, as $\tau_{dust}\rightarrow1$, any absorption signature from the molecular line approaches zero.
This means we should not expect to see the highest velocities in absorption at all, but we should see very strong absorption from cooler gas further in front of the core.
The point at which we begin to detect molecular absorption indicates the approximate location of the $\tau_{dust}=1$ surface, but it is not a very strong probe since what we observe is a mass-weighted profile of the foreground gas.
This mechanism readily explains the lack of signatures in the absorption lines toward each source. 
\item
{\it  A purely observational effect hides the emission. } 
When lines appear as absorption features toward the continuum source and emission features off of the source, the average surface brightness in a given velocity channel approaches a smooth, constant value.
Because we are observing with an interferometer with low spatial dynamic range, these smoother features are suppressed.
It is likely that this effect is sufficient to push any emission signatures below the noise of our current data set.
\end{enumerate}

We regard explanation (2) above as the most relevant.  It implies that longer-wavelength observations, where the dust may be optically thin, are necessary to detect infall kinematics and to measure dynamical masses  of high-mass protostars.
 We note that our finding that that dust opacity  prevents  kinematic measurements on small scales in high-mass protostellar disks   using dense gas molecular lines was explicitly predicted in \cite{Krumholz2007}. 
 
\section{Spatial and velocity structure of outflows}
\label{app:outflows}
Here, we complement high-angular resolution images of the thermal SiO $J=5-4$ line from ALMA cycle 3 (beamsize 0\pas035) with lower-resolution images of \co\  $J=2-1$ from ALMA cycle 2 (beamsize 0\pas2) as well as VLBA measurements of \wat\ masers (beamsize 0\pas001). 
We focus on W51-e2e and W51-North because their outflows display complex spatial and velocity structures (we were unable to trace the larger scale outflow from W51-e8).
The combination of different tracers provides us with a full picture of these outflows,  on scales ranging from hundreds  up to tens of thousands of AU.

\subsection{W51-e2e}

W51-e2e drives a prominent bipolar outflow, which was first detected in the CO (3-2) line with the SMA by \citet{Shi2010b}, and then imaged with ALMA in the CO (2-1) line at 0\pas2 resolution by \citet{Ginsburg2017} (Fig.~\ref{fig:outflow_e2e},  left panel). 
This outflow has a high relative velocity $v \pm 100$ \kms.   
On the largest-scales traced by \co, both ends (red- and blue-shifted) of the outflow are sharply truncated at $\sim$ 2\pas5 (0.07 pc or 15000~AU). 
To the  southeast (SE), the high-velocity flow lies along a line that is consistent with the extrapolation from the
 NW flow (i.e. directly bi-polar), however at lower velocities ($10 < V_{LSR} < 45$ \kms) the direction appears more to the south, before being abruptly truncated.   
 For the NW redshifted part of this outflow ($70 < V_{LSR} < 120$ \kms) there is a clear bend apparently at an intersection with a blueshifted flow ($22 < V_{LSR} < 45$ \kms) from another source, W51e2nw, that is also believed to be an intermediate-to-high-mass protostar \citep{Goddi2016}. The significant change in direction suggests that these two outflows interact. Although such a scenario seems implausible given the outflows small volume filling factor, both the morphology and the distinct velocity range of the two outflows (which makes it easy to distinguish the two) point to a genuine encounter. Therefore, at least for the NW flow, this encounter is responsible for the truncation of the redshifted lobe.  
 
 The ALMA long-baseline images unveil the smallest scales of this outflow (Figure~\ref{fig:outflow_e2e}, right panel).
  Although the SiO structure is broadly consistent with the lower-resolution \co\ structure at the base of the outflow, the P.A. of the line connecting the  NW and  SE flows is --56\dg\ (whereas the \co\ outflow is closer to P.A.=--37\dg).  
The blue-shifted flow extends only across $\sim$0\pas45 or 2430~AU.
 The bulk emission of the redshifted lobe in the NW has a similarly compact size ($\sim$0\pas6 or 3240~AU), but some weaker emission  extends across $\sim$10000~AU and is co-spatial with the \co\,outflow  (this more extended, weaker structure is  recovered in lower-resolution images of SiO tapered to 0\pas1, not displayed here). 
 Interestingly, the redshifted lobe  shows a two-stream limb-brightened morphology with a central cavity, possibly due to the presence of a dusty filament at the center (see left panel of Fig.~\ref{fig:rgb_overlays}). 
 The dynamical age of the SiO outflow is $\sim$116 years (see Appendix~\ref{app:outflow_age}), shorter than the 600 years estimated by \citet{Ginsburg2017} at the peak observed velocity of \co.

\citet{Sato2010} used the VLBA to measure proper motions  of water masers, which also show a picture consistent with the thermal \co\ and SiO emission. 
Interestingly, the blueshifted and redshifted water masers closer to the protostar imply a flow P.A. of --56\dg, perfectly consistent with thermal SiO, while the redshifted masers along the outflow at larger distances, have a P.A. of --43\dg, closer  to that of the \co\,outflow axis spatially distant from the protostar. Considered together, these findings are consistent with an indication of change in P.A. of the outflow with increasing distance from the protostar.

\subsection{W51-North}

The outflow from W51-North is remarkably extended and complex. 
Figure~\ref{fig:outflow_n} offers a full view of this outflow, on scales from 25000~AU down to hundreds of AU. 
The first notable feature is that the redshifted and blueshifted  lobes on the large-scales as traced by \co\ are significantly asymmetric. 
A collimated high-velocity component (covered velocities of $\pm$100 \kms) extends directly north across$\sim$4\arcsec\ (or  20000 AU), while   the blueshifted component points to the  SE and it is sharply truncated, extending only across $\sim$0\pas6 (or  3200 AU)  (Fig.~\ref{fig:outflow_n}, upper panel;  \citealt{Ginsburg2017}). At the smallest scales, i.e., within a few thousands AU from the driving source (Fig.~\ref{fig:outflow_n}, lower panel), 
further surprises are revealed. 
 At the base of the \co\ outflow,  SiO reveals  a very fast and well collimated bipolar outflow, with the blue lobe extending across 0\pas34 (or 1600~AU) and having a maximum speed of 96 \kms. The red lobe is slightly more compact with a radius of 0\pas24 or 1300~AU and a maximum speed of 66~\kms.  
 Both SiO lobes are consistently oriented at P.A. $\sim -$70\dg, clearly rotated from the large-scale \co\ which has P.A. -13\dg (Fig.~\ref{fig:outflow_n}, upper panel). 

Close to the central protostar, the  redshifted  outflow lobe shows  a remarkable structure. It has an orientation consistent with that of the blueshifted lobe at a position angle (P.A).$\sim -$70\dg, up to a radius of 0\pas24 (1300~au). 
 Starting at about 0\pas15 (800~au) from the center, it suddenly turns by 80\dg (at P.A.$\sim$10\dg) and curves until it settles at P.A. 0\dg\ along the base of the redshifted lobe of the \co\,outflow. 
The high-velocity portions of the flow (up to $\sim$70 \kms from  the systemic velocity) are found closer to the protostar, while the material in the `transition' region has lower-velocity ($65 < V_{LSR} < 90$ \kms, i.e. within 30 \kms from the stellar systemic velocity). Some  high-velocity material (up to 50 \kms from  the systemic velocity) is also observed at the northern end of the SiO emission (see Appendix~\ref{app:vel_outflows} and Figs.~\ref{figapp:outflow_e2e} and~\ref{figapp:outflow_north} for a description of the velocity structure of the  outflows). 
The large-scale red lobe shows contiguous structure tracing onto the small-scale, with the change in direction starting at around 
0\pas35 (1900~au) and completing by about 0\pas15 (800~au).
Assuming a constant velocity of 30 \kms, the outflow turn occurred over just  175 years.
The general structure, including the sharp turn seen in the redshifted lobe, is reflected in the blueshifted lobe, which shows a southward `protrusion' starting at about 0\pas22 (1200~au) from the protostar. 
Although much less prominent, this   southern component in the blueshifted SiO lobe provides a (weaker and more compact) counterpart to the redshifted northern component.
In both SiO and CO emission, the blueshifted lobe is barely detected beyond about 0\pas30 or 1600 au (although the weakest CO emission extends up to 3200~AU); one possibility is that it breaks out of the cloud (assuming the source is located at the front of it). 

The peculiar structure of the outflow revealed by ALMA images of SiO emission is also tracked by the \wat\ masers 3-D velocities \citep{Imai2002}, shown as colored arrows in the right panel of Fig.~\ref{fig:outflow_n}. The \wat\ masers reside in two complexes separated by $\sim$3000~AU at the front edges of the compact SiO outflow lobes\footnote{ We registered the \wat\,masers positions and our ALMA maps under the assumption that the \wat\,maser outflow origin (estimated with a least-squares fitting analysis of the measured maser 3-D motions - \citealt{Imai2002}) 
is coincident with the peak of the dust continuum and/or the origin of the SiO outflow imaged with ALMA. We assume that this is the putative location of the driving protostar. } 
 and are moving away from each other in ballistic motions at about 200~\kms\  and at a position angle of~--72\dg, consistent in both P.A. and velocity with the SiO compact outflow. 
In particular, the \wat\ masers detected in the  SE appear to trace a bow shock at the tip of the blueshifted SiO lobe, while the \wat\ maser complex in the opposite side traces both the high-velocity  NW redshifted lobe as well as the base of the northern outflow component.
In the redshifted part of the outflow, the measured proper motions appear to smoothly rotate from  NW to north moving away from  the tip of the SiO lobe, following closely the SiO emission. 
In addition, the expansion velocity of the outflow decreases from 90~\kms\ to 50~\kms\ at radii 0\pas2 to 0\pas5   from the outflow origin, also consistent with the SiO redshifted emission. 
Finally, a least-squares fitting analysis to the measured maser 3-D motions  provides evidence of a tri-axial symmetry, inconsistent with a simple bipolar symmetric outflow, but consistent with multiple outflows or with a single outflow with a more complex structure. 

\subsection{Velocity structure as a function of distance}
\label{app:vel_outflows}

\begin{figure*}
\centering
\includegraphics[width=0.4115\textwidth,trim={0cm 0cm 2.47cm 0cm},clip]{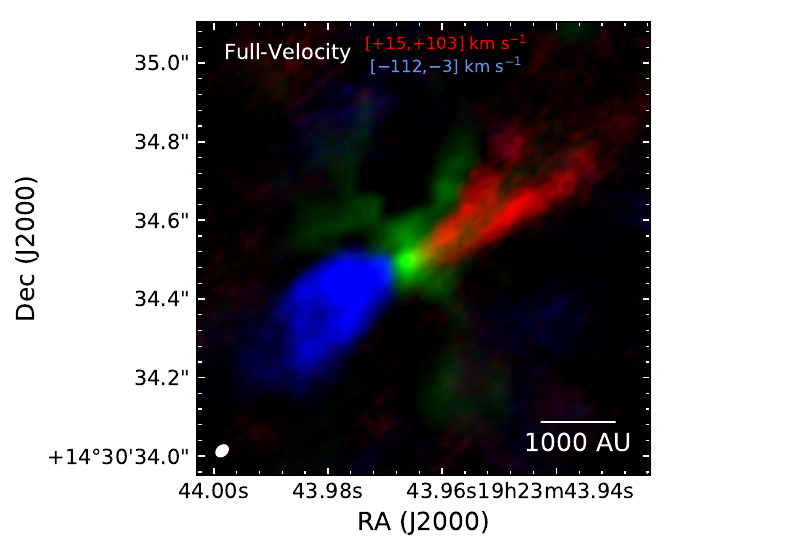} 
\includegraphics[width=0.29\textwidth,trim={2.35cm 0cm 2.47cm 0cm},clip]{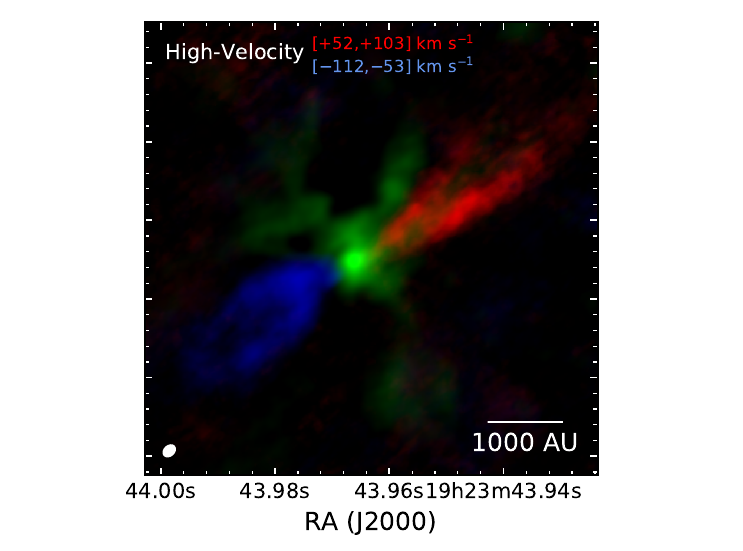}
\includegraphics[width=0.29\textwidth,trim={2.35cm 0cm 2.47cm 0cm},clip]{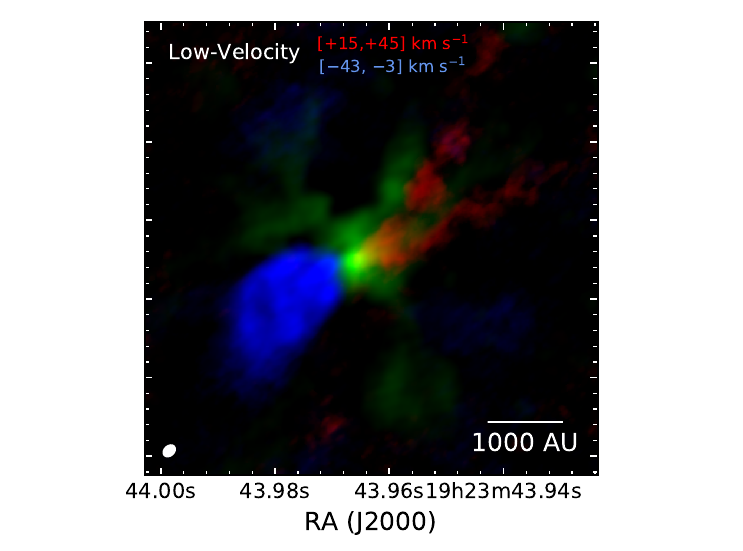}
\caption{Structure of the massive protostellar outflow driven by W51e2e as a function of velocity within 3000~AU from the central protostar.  
The red and blue colors indicate redshifted and blueshifted emission of  the SiO J = 5-4 line. The emission is integrated over different velocity ranges (indicated in brackets at the top of each panel): full-velocity range (left panel), high-velocity range (middle panel), low-velocity range (right panel). 
The velocity ranges are with respect to the systemic velocity of the protostar (56 \kms). 
The green colour displays the dust continuum emission at 1.3~mm, already shown in Figure 1 in the main text; here a different 'stretch' of the intensity brightness   ({\it arcsinh} function in the {\it matplotlib} Python plotting library) is employed to highlight the central core (and filter out the complex filamentary structure).  
The angular resolution of these data is given by the ellipse drawn in the lower left corners (same as figure 1 in the main text).  
}
\label{figapp:outflow_e2e}
\end{figure*}

\begin{figure*}
\centering
\includegraphics[width=0.414\textwidth,trim={0cm 0cm 2.42cm 0cm},clip]{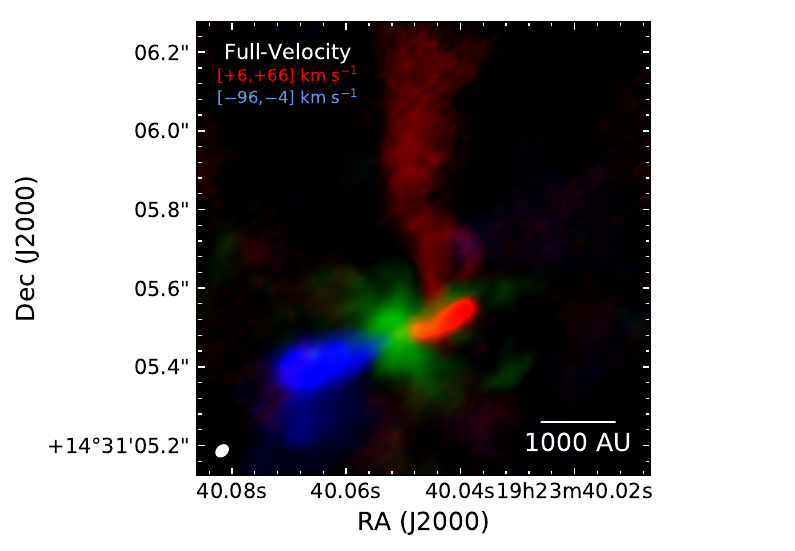} 
\includegraphics[width=0.288\textwidth,trim={2.46cm 0cm 2.42cm 0cm},clip]{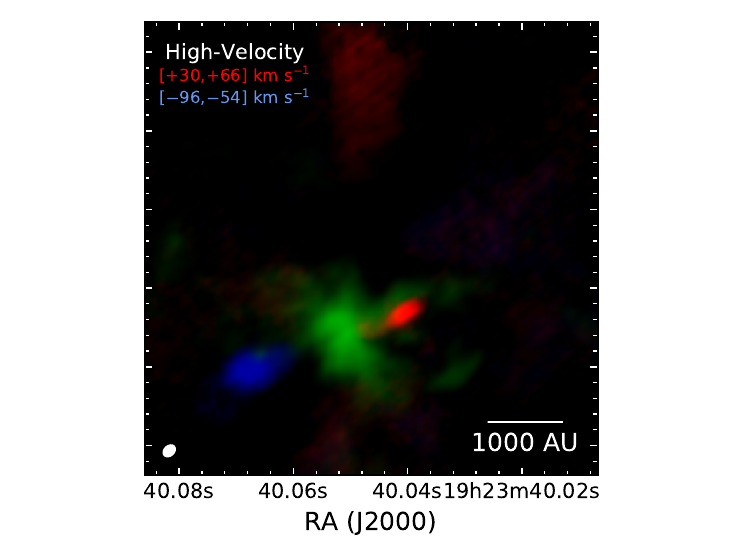}
\includegraphics[width=0.288\textwidth,trim={2.46cm 0cm 2.42cm 0cm},clip]{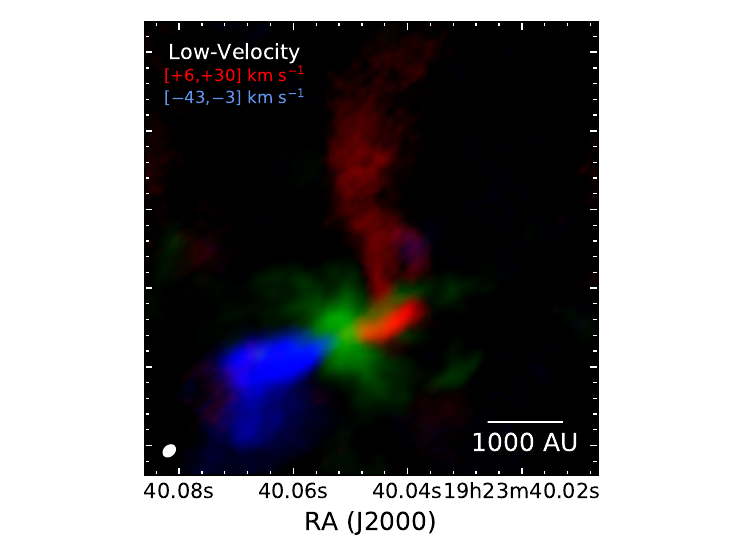}
\caption{Structure of the massive protostellar outflow driven by W51north as a function of velocity within 4000~AU from the central protostar.  
The red and blue colors indicate redshifted and blueshifted emission of  the SiO J = 5-4 line. The emission is integrated over different velocity ranges (indicated in brackets in the upper left corner of each panel): full-velocity range (left panel), high-velocity range (middle panel), low-velocity range (right panel). 
The velocity ranges are with respect to the systemic velocity of the protostar (60 \kms). 
The green colour displays the dust continuum emission at 1.3~mm, already shown in Figure 1 in the main text; here a different 'stretch' of the intensity brightness   ({\it arcsinh} function in the {\it matplotlib} Python plotting library) is employed to highlight the central core (and filter out the complex filamentary structure).  
The angular resolution of these data is given by the ellipse drawn in the lower left corners (same as figure 1 in the main text).  
}
\label{figapp:outflow_north}
\end{figure*}

Figures~\ref{figapp:outflow_e2e} and~\ref{figapp:outflow_north} showcase the structure of the SiO protostellar outflows as a function of velocity within 2000 au from the central protostars W51e2e (top two rows) and W51north (bottom two rows), respectively. 
Note that in W51e2e the high velocities trace preferentially the central parts (i.e., primary wind) and the low-velocities the limb-brightened edges (i.e., entrained gas) of the outflow, a phenomenon often observed in low-mass protostellar outflows  \citep{Frank2014}. In W51north the high velocities trace material at larger radii from the protostar both in the compact  NW- SE outflow and the larger scale north-south outflow, reminiscent of the `Hubble flows'  observed in low-mass protostellar outflows.  
It is also interesting to note that the bulk of the compact  NW- SE outflow  expands at high velocities while  the bulk of the larger scale north-south outflow expands at low velocities, in agreement with the scenario where the former represents the latest outflow event and the latter is a fossil outflow. 

\section{Physical properties of massive protostellar outflows}
\label{app:outflow_properties}

\subsection{Dynamical ages}
\label{app:outflow_age}

To estimate the dynamical age of the outflows, $T_{dyn}$, we take the ratio of the  projected  length (measured from the protostar), $R_{max}$, to the the maximum speed of the outflow, $V_{max}$, corrected for the inclination $i$: $T_{dyn} (V_{max}) = ( R_{max} / V_{max} ) / \tan{(i)}$\footnote{This method is adequate  since the highest speeds are observed at the largest separations (see Appendix~\ref{app:vel_outflows}).}. 

Toward W51e2e, the outflow is clearly bipolar and fairly symmetric, though the redshifted side is partly obscured by the dust continuum, therefore   we use the blue lobe in our estimates.
We observe a maximum velocity of 105 \kms ($v_{lsr}=-50$ \kms) at a separation of about 2560 AU, which gives a dynamical age  $T_{dyn} (V_{max}) = 116 \times (\tan{(45^{\circ})}/\tan{(i)})$ years, where the inclination $i$ is expressed in units of degrees.  

In W51e8, the highest observed  outflow velocity is 42 \kms, at a separation of about 390~AU, which corresponds to $T_{dyn} (V_{max}) = 44 / \tan{(i/ 45^{\circ})}$ years. 
We however warn the reader that  the outflow appears to be nearly in the plane of the sky,  therefore the age estimate is not reliable.

In W51north, we first consider the blueshifted lobe, which displays a much simpler structure. 
We observe a maximum velocity of 95 \kms\ ($v_{lsr}=-35$ \kms) at a separation of about 1540 AU, which gives a dynamical age  $T_{dyn} (V_{max}) = 77 \times (\tan{(45^{\circ})}/\tan{(i)})$ years. 

The redshifted  lobe from W51north breaks in three major components, representing three consecutive events: these are labeled as 'young', 'transition', and 'old' in  Fig.~\ref{fig:outflow_n_sketch}. 
The 'young' flow is the component along NW- SE close to the protostar. 
For a maximum velocity of 66 \kms\  at a separation of about 1300 AU, the dynamical age is $ \sim 100 \times (\tan{(45^{\circ})}/\tan{(i)})$ years, higher than inferred from the blueshifted lobe. Adopting an average velocity of $\sim$30~\kms\ at a separation of 800~AU (where the second component starts), one gets 125 years.
The second component, which identifies  the "transition" between the young  NW- SE component and the old north-south component, extends from 800 AU to 1900~AU, corresponding to 300 to 125 years (or a duration of 175 years) using an average velocity of $\sim$30~\kms.  
The third component is the larger-scale "old" north-south outflow
and extends up to 0\pas8 or about 4300 AU, yielding  an age  of 680 years for a $\sim$30~\kms\ velocity.  
An even older component of the outflow is traced by  \co, up to a length of about 4~arcseconds or 22000~AU, corresponding to an age of about 3500 years (for an average velocity of $\sim$30~\kms).

\subsection{Momentum rate}
\label{app:momentum-rate}

In order to determine the momentum of the outflowing gas, 
we  compute  its total mass from the SiO emission integrated in the full velocity range. 
We follow a standard approach in two steps. 
 
As a first step, we measure the number of SiO molecules per unit area along the line of sight. 
We first calculate the column density of  SiO in the energy level corresponding to the observed transition J=5-4, using the measured intensity of that transition (channel by channel) and the Boltzmann equation for statistical equilibrium coupled with the standard radiative transfer equation -- e.g. the optically thin version of Eq. (30) in \cite{MangumShirley2015}.  
Then we relate the number of molecules in the given energy level 
to the total population of all energy levels in the molecule assuming that they are populated according to a Boltzmann distribution: 
we use the “rotational partition function”, a quantity that represents a statistical sum over all rotational energy levels in the molecule, 
and we assume a constant temperature defined by the excitation temperature $T_{ex}$ -- e.g. Eq. (31) in \cite{MangumShirley2015}. 
We adopt $T_{ex}=$250~K (note that the partition function is approximately linear with $T_{ex}$ in this regime, so if the molecules are twice as hot, the column estimate should be halved). 

In the second step, we use the  SiO column density to compute the total mass of the molecular gas.  
To convert from SiO column to \hh\ column, we assume a conservatively high abundance of SiO, $N(SiO) / N(\hh) = 10^{-7}$. 
The total momentum is then obtained by integrating over the full velocity range of SiO emission: ${P}_o = \sum_v  m_v \ v$.

In our analysis, we treated the blueshifted and  the redshifted  lobes independently, 
but the latter provided mass values 4 to 10 times lower. 
In fact, in W51e2e, the redshifted lobe is partly obscured by the dust continuum, whereas in W51north the redshifted side shows a more complex structure than the blueshifted side, yielding more uncertain estimates. To mitigate these observational biases, we used the blue flows alone in our estimates.  
The outflow masses and momenta derived with this analysis are reported in columns 6 and 7 of Table~\ref{tab:outflows}.
In order to account for both outflow lobes we thus need to multiply those numbers by a factor of 2, yielding   total masses of 0.36, 0.72, 0.22~\ms\ and total momenta of 18, 50, 3~\ms~\kms, for the outflows driven by W51north, W51e2e, W51e8, respectively.

\section{Disk angular momentum}
\label{app:diskL}

We define the total angular momentum in a Keplerian disk about a protostar with mass $M_*$ as: 
$L_{disk} = \int_{0}^{R} {M(r) V(r) dr}$ where: 
$V(r)=\sqrt{GM_*/r} \ {\rm and} \ M(r) = 4\pi r^2 \rho$. 
 We assume for simplicity that the gas density $\rho$ is constant as a function of radius\footnote{This is a conservative assumption since a power-law  $\rho\propto r^{-\alpha}$ would give lower $L$ for positive $\alpha$, resulting in an even greater flipping (following a substantial dump mass onto the disk) for any realistic disk with  $\alpha=1$ or  $\alpha=2$.}: $\rho = M_{disk}/4\pi R^2_{disk}$. Therefore: 
 $L_{disk} = \int_{0}^{R_{disk}} G^{0.5} M_*^{0.5} (4\pi \rho) r^{3/2} dr = (2/5) 4\pi G^{0.5} M_*^{0.5} \rho  R_{disk}^{5/2}  = 0.4 M_{disk} \sqrt{G  M_*    R_{disk}}$. 
If we assume $R_{disk} = 350$~AU, $M_* =20$~\ms, and $M_{disk} = 1$~\ms, then  $L_{disk} = 3 \times 10^{54}$~g~cm$^2$~s$^{-1}$.

\section{Disks around HMYSOs known from previous ALMA studies}
\label{app:HMdisks}

\input{table_HMdisks}

In Table~\ref{tab:HMdisks} we report the physical properties of known Keplerian disk candidates around HMYSOs  observed with ALMA. 
The list includes seven B-type HMYSOs and seven O-type  HMYSOs where  Keplerian rotation signatures have been found  to date with ALMA observations.
Typically, these studies analyze the velocity field of the circumstellar gas as revealed by the line emission of a high-density molecular tracer (e.g., CH$_3$CN) and base 
their claims on two properties: 
(1) a velocity gradient along the major axis of the source in $^{1st}$moment maps; 
(2) a clear a “butterfly" pattern with 
high-velocity spikes in correspondence of the  HMYSO position  in position-velocity plots. 

Table~\ref{tab:HMdisks} is an update of Table 1 in \citet{Rosen2020review}.  
\citet{Johnston2020} and \citet{Beltran2020} report independently  a slightly different list (thirteen each)  of  disk candidates around O-type HMYSOs. Those lists exclude some of the B-type HMYSOs listed in Table~\ref{tab:HMdisks}, but include disk candidates around O-type HMYSOs without Keplerian profile signatures.

%% file: table_HMdisks.tex
\begin{deluxetable*}{lcccccr}
    \tablecaption{Properties of Keplerian disk candidates around HMYSOs as observed with ALMA to date. }
\tablewidth{0pt}
\tablehead{ 
\colhead{Object} & \colhead{Distance} & \colhead{Luminosity} & \colhead{Star Mass} & \colhead{Disk Mass} & \colhead{Disk Radius}  & \colhead{References} \\
         & \colhead{[ kpc ]} & \colhead{[ \ls ]} & \colhead{[ \ms ]}    & \colhead{[ \ms ]}    & \colhead{[ AU ] } &
         }
\startdata
\noalign{\smallskip}
\multicolumn{7}{c}{B-type YSOs} \\    
\noalign{\smallskip}
  Orion Source I                  & 0.4 & $\sim10^4$ & $15\pm2$ & $<0.2$ & 75-100  &  [1,2] \\
   IRAS~20126+4104          & 1.6 & $\sim10^4$ & 12 & 1.5 & 860 &  [3,4] \\
   IRAS~18162--2048\tablenotemark{a}  & 1.7 & $\sim10^4$ & 18 & 4 & 300  & [5] \\
   G339.88-1.26                 & 2.1 & $4 \times 10^4$ & $11\pm5$ & --- & 430--630 & [6] \\
   G35.20-0.74N                  & 2.2 & $\sim10^4$ & 18 $\pm$ 3 & 3 & 2500 & [7] \\ 
   G35.03+0.35 A                 & 3.2 & $6 \times 10^3$ & $9\pm4$ & 0.75 & 2200 & [8,9]  \\
   G16.59$-$0.05                 & 3.6 & $3 \times 10^4$ & $10\pm2$ & $1.8\pm0.3$ & 500 & [10]  \\
\noalign{\smallskip}
\hline 
\noalign{\smallskip} 
\multicolumn{7}{c}{O-type YSOs}  \\   
\noalign{\smallskip}   
   S255IR NIRS3              & 1.8 & $1.6 \times 10^5$\tablenotemark{b} & 20 & 0.3 & 500 & [11] \\ 
   G351.77-0.54                & 2.2\tablenotemark{c} & $1.7 \times 10^4$    & 14-25\tablenotemark{d} & 0.1-0.49\tablenotemark{e} & 250-500\tablenotemark{b} & [12] \\
   G17.64+0.16                 & 2.2 & $\sim 10^5$  & $45 \pm 10$ & $<2.6$ & 120 & [13,14] \\
   IRAS16547-4247          & 2.9 & $\sim 10^5$   & 20 & 4 & 870 & [15,16] \\
    G11.92-0.61 MM1       & 3.4 & $\sim 10^4$\tablenotemark{f} & $34\pm5$ & 2.2-5.8 & 480 & [17] \\
    AFGL 4176                 & 4.2 & $\sim  10^5$ & 20 & 2--8 & 1000 & [18,19]\\
    G023.01$-$00.41       & 4.6 & $4 \times 10^4$  & 20  & 1.6 & 2500 & [20] \\
\enddata
\tablecomments{ When mass error bars are not given, the measurements should be taken as loose estimates, e.g., based on consistency checks between a stellar type and the upper-limit luminosity. 
Radius estimates are wavelength-dependent. 
 \tablenotetext{a}{Also known as GGD27 MM1, the protostar driving the HH80/81 jet.}
\tablenotetext{b}{Source  luminosity  increased from $2.9 \times 10^4$ to $1.6 \times 10^5$ as a consequence of an accretion burst recorded on Nov 2015.}
\tablenotetext{c}{The distance to the source is either 1 kpc or 2.2 kpc.}
\tablenotetext{d}{Estimated from the source bolometric luminosity through simulated stellar cluster \citep{BdW16}.}
\tablenotetext{e}{Depending on the distance to the source assumed. }
\tablenotetext{f}{The low luminosity in a $>30$~\ms\ YSO could be explained either with the presence of a binary (e.g., two $\sim$15~\ms\ YSOs) or with a high accretion rate, which would increase the protostellar radius and decrease its effective temperature \citep[e.g.][]{Hosokawa2009}.}
}
\tablerefs{[1,2]: \citet{Plambeck2016,Ginsburg2018}; 
[3,4]: \citet{Cesaroni2014,Chen2016}; [5]:  \citet{Girart2018}; [6]: \citet{Zhang2019}; [7]: \citet{Sanchez-Monge2013a}; [8,9]: \citet{Beltran2014,BdW16}; [10]: \citet{Moscadelli2019}; [11]: \citet{Caratti2017};  [12]: \citet{Beuther2017}; [13,14]:  \citet{Maud2018,Maud2019}; [15,16]: \citet{Zapata2019,Tanaka2020}; [17]: \citet{Ilee2018}; [18,19]: \citet{Johnston2015,Johnston2020}; [20]:  \citet{Sanna2019}. }
  \label{tab:HMdisks}
 \end{deluxetable*}